\documentclass[10pt]{article}%
\usepackage[utf8]{inputenc}%
\usepackage[english]{babel}%
\usepackage[dvips]{graphicx}%
\usepackage{hhline}%
\usepackage{longtable}%
\renewcommand{\thefootnote}{\ifcase\value{footnote}\or(*)\or
(**)\or(***)\or(****)\fi}
\begin{document}

\twocolumn[\protect{%
\begin{center}
{\bf \large GALAXY MOTIONS IN THE BOOTES STRIP}\\
\bigskip
{\large{}I.\,D.\,Karachentsev$^{1,\,2\,}$\footnotemark,
V.\,E.\,Karachentseva$^{3\,}$\footnotemark,
O.\,G.\,Nasonova$^{1\,}$\footnotemark\\}
\bigskip
{\footnotesize\itshape$^{1}$Special Astrophysical Observatory of RAS, Nizhnij Arkhyz, Russia\\
$^{2}$Leibniz-Institut f\"{u}r Astrophysik, Potsdam, Germany\\
$^{3}$Main Astronomical Observatory of NASU, Kyiv, Ukraine\\}
\bigskip
\end{center}

\begin{quote}
We explore the structure and kinematics of a dispersed filament of galaxies
residing between the Local Void and the Virgo cluster. For such purpose, we
consider a sample of 361 galaxies with radial velocities $V_{LG} <
2000$~km$\:$s$^{-1}$ inside the sky area of [RA$=13.0^h-18.0^h$ and
Dec.$=-5^{\circ} - + 10^{\circ}$]. At present, 161 of them have individual
distance estimates. The galaxy distribution on peculiar velocities along the
strip exhibits the known Virgo-centric infall at $RA < 14^h$ and some signs of
outflow from the Local Void at $RA > 17^h$. Majority of the Bootes strip
galaxies (56\%) belong to 13 groups and 11 pairs, with the most prominent group
around NGC\,5846. The Bootes strip groups reside within [$17-27$]~Mpc, being all 
farther from us than the Virgo cluster. The Bootes filament contains the total
stellar mass of $2.7\times10^{12} M_\odot$ and the total virial mass of 
$9.07\times10^{13} M_\odot$, having the average density of dark matter to be
$\Omega_m = 0.09$, i.e. a factor three lower than the global cosmic average.
\end{quote}

{\bf{}Keywords}: galaxies: distances and redshifts, (cosmology:) large-scale structure of Universe}

\vspace{1cm}]

\footnotetext[1]{E-mail: ikar@sao.ru}
\footnotetext[2]{E-mail: vakara@mao.kiev.ua}
\footnotetext[3]{E-mail: phiruzi@gmail.com}

\section{Introduction}

Recent mass measurements of radial velocities and distances of galaxies  
engage for obtaining shortly the detailed map of peculiar motions
within the Local Supercluster. Since the peculiar velocities field is generated 
by the dark matter distribution
in the considered volume and its neighbourhood, invoking N-body simulations to
the analysis gives us a possibility to trace the large scale structure relief,
i.e. the disposition of the main attractors and voids. According to Sorce et al.
(2013), the constrained simulations method can spot some massive proximal
attractors such as the Virgo cluster with an accuracy of about 5 Mpc.

However, as it was noticed by different authors (Vennik 1984, Tully 1987,
Makarov \& Karachentsev 2011, Karachentsev 2012), the virial mass estimates of
galaxy groups and clusters in the Local Universe lead to the average matter
density value of $\Omega_m(local)\simeq 0.08$ which is 3 times lower than the
global cosmological value $\Omega_m(global)= 0.24\pm0.03$ (Spergel et al. 2007).
One of the possible explanations of this discrepancy is the assumption that the
lacking 2/3 of the dark matter total amount are spread outside the virial radii
of galaxy groups and clusters. Yet, the analysis of Hubble flows around the
nearest aggregates: the Local Group (Karachentsev et al. 2009), M81
group (Karachentsev \& Kashibadze 2006), Cen~A group (Karachentsev et al. 2007),
Virgo cluster (Karachentsev \& Nasonova 2010) and Fornax cluster (Nasonova et
al. 2011) shows that the total masses of these groups and clusters within ``zero
velocity surface'' radii, $R_0$, are in a good agreement with the virial mass
estimates, although $R_0$ radius is roughly 4 times larger than virial one.
Hence, the major fraction of the lacking dark matter is spread outwith infall zones
around groups and clusters. There are some suggestions in literature that
significant amount of dark matter could be located in ``dark filaments''
conducting intergalactic matter into hot virial regions (Dietrich et al. 2012,
Whitting 2006).

Obviously, the Local Supercluster is the most appropriate object to verify these
hypotheses due to the high density of observational data on galaxy
velocities and distances. In our previous papers we considered motions of
galaxies in filaments attached to the Virgo cluster, as the centre of the Local
Supercluster, from North and South. In Virgo Southern Extension region
[RA=$12.5^h-13.5^h$, Dec. =$-20^{\circ} - 0^{\circ}$] (Karachentsev \& Nasonova
2013) and Ursa Majoris region [RA=$11.0^h-13.0^h$, Dec. =$+40^{\circ}-
+60^{\circ}$] (Karachentsev et al. 2013) the mean matter density was estimated
to be $\Omega_m=0.11$ and 0.08, respectively. But we surmised the existence of a
dark attractor with the mass of $\sim2\times 10^{14}M_{\odot}$ in the
Coma~I region [RA=$11.5^h-13.0^h$, Dec. =$+20^{\circ} - +40^{\circ}$]
(Karachentsev et al. 2011). All these three zones are situated along the equator
of the Local Supercluster where the structure and kinematics of galaxy groups is
some way faded away due to projection effects.

As it can be seen from the mapped distribution of galaxies with radial
velocities $V_{LG}<2000$~km$\:$s$^{-1}$ (Figure~1), there is a chain of galaxy
groups spanning from Virgo [RA=$12.5^h$, Dec. =$+12^{\circ}$] to the Local Void
direction [RA=$19.0^h$, Dec. =$+3^{\circ}$]. The kinematics of this structure
should be influenced both by galaxies infalling towards the Virgo cluster as
well as more eastern galaxies moving away from the expanding Local Void
(Nasonova \& Karachentsev 2011). This narrow strip of sky crossing the Bootes
constellation was decided to be considered in details in the present paper.

\section{Observational data on galaxies in the Bootes strip}

Based on the Lyon Extragalactic Database = LEDA (http://leda.univ-lyon1.fr), 
we have selected galaxies with radial velocities $V_{LG}\leq 2000$~km$\:$s$^{-1}$
relating to the Local Group centroid in the sky region with equatorial
coordinates RA = [$13.0^h, 18.0^h$], Dec. = [$-5^{\circ}, +10^{\circ}$]. Among 2515
extracted objects, 2154 or 86\% (!) turned out to be false galaxies, mainly the
Milky Way stars. There is also a number of high velocity clouds detected in
Arecibo HI survey.

The major part of the considered strip is covered by the optical SDSS survey
(Abazajian et al. 2009) as well as HIPASS (Zwan et al. 2003) and ALFALFA (Haynes
et al. 2011) HI surveys, which is a certain advantage for our purposes. Using
these surveys we performed an independent morphological classification of
galaxies and refined the data on their apparent magnitudes and radial
velocities. We eliminated several cases when apparent magnitude or radial
velocity estimates were attributed to fragments of the same galaxy. New data
from the ALFALFA survey gave us an opportunity to determine distances for many
galaxies from the Tully \& Fisher (1977) relation between luminosity of a galaxy
and its HI line width $W_{50}$ (measured at the level of 50\% of the peak).

The resulting list of 361 galaxies in Bootes strip obtained by comparing
critically the data from different sources and eliminating some ambiguous cases
is presented in Table 1. Its columns contain: (1) galaxy number in the known
catalogues; (2) equatorial coordinates for 2000.0 epoch; (3) integral apparent
magnitude in the $B$ band from the NASA Extragalactic Database (=NED)
(http://ned.ipac.caltech.edu), LEDA or SDSS; in some cases with discrepant
values of $B_T$ from different sources we had reliance on our own eye estimates
of the apparent magnitude; (4) galaxy distance (in Mpc) together with the method
applied for estimating the distance:  ``sn''~-- from Supernova luminosity,
``rgb''~-- from the tip of the red giant branch luminosity, ``sbf''~-- from 
surface brightness fluctuations (Tonry et al. 2001), ``tf'' or ``TF''~-- from the
Tully-Fisher relation; in most cases (indicated as TF) distances were estimated by us
independently using the relation from Tully et al. (2009):

$$M_B=-7.27(\log W_{50}-2.5) -19.00;$$

to correct $W_{50}$ for inclination in the case of dwarf galaxies we assumed 
their spatial shape to be a spheroid with the axial ratio of 1:2 (Roychowdhyry 
et al. 2013); (5) radial velocity in the Local Group frame and its error (in 
km$\:$s$^{-1}$); (6) morphological type according to our identification; (7) name
of the brightest galaxy in the group (Makarov \& Karachentsev 2011) or in the
pair (Karachentsev \& Makarov 2008) which the considered galaxy belongs to. As
it follows from this column data, more than a half (56\%) of all galaxies in the
Bootes strip form bound systems.

The upper panel of Figure~2 presents the radial velocity distribution of 361
galaxies in the strip. The velocity colorscale is shown under the panel. All
the galaxies have radial velocities $V_{LG} > 650$ km$\:$s$^{-1}$ except proximate
dwarf KKH~86 with $V_{LG} = 209$ km$\:$s$^{-1}$ and $D$= 2.6 Mpc. Most galaxies
with radial velocities lying in the range (700~-- 1300 km$\:$s$^{-1}$) are situated
on the western side of the strip, neighbouring the Virgo cluster. A circular arc
near RA~$\simeq 14^h$ marks the zero velocity surface with radius $R_0$ which
separates galaxies falling toward the Virgo centre from those being involved with
the cosmological expansion. For the Virgo cluster $R_0$ value is 7.2 Mpc or
25$^{\circ}$ (Karachentsev et al. 2014). MK groups (Makarov \& Karachentsev
2011) are marked by names of their brightest galaxies. The most notable feature
of galaxy distribution in the Bootes strip is the compact group NGC\,5846 which
numbers 74 members with measured radial velocities (some of them slightly exceed 
the limit of 2000~km$\:$s$^{-1}$ adopted for this research).

The middle panel of Figure~2 represents the morphological type distribution of
considered galaxies. Early type galaxies (E, S0, dSph) are plotted as red
markers, spirals (Sa--Sdm) as green ones while irregular galaxies (Ir, Im) and
blue compact dwarves (BCD) marked with blue. As one can see, most early type
galaxies are concentrated among the population of NGC\,5846 group and several
other groups (NGC\,5363, NGC\,5638), though certain S0 galaxies (NGC\,6010, 
CGCG~052-15) occur in the general field too.

Among all the 361 galaxies of the Bootes strip, 161 galaxy (45\%) have distance
estimates. Distance distribution of these galaxies is shown on the lower panel
of Figure~2. About 2/3 of them lie in the range of (25$\pm$5) Mpc. As most
galaxies in the Bootes strip have distances measured from Tully-Fisher relation
with an accuracy of $\sim20$\%, the line-of-sight width of the Bootes filament
turns out to be comparable with the typical distance error. A polyline under the
panel represents the running median of $D$ value along right ascension with
window of $0^h\hspace{-0.4em}.\,5$. As it follows from these data, the major part of the Bootes
filament galaxies is located farther than the Virgo cluster which has, according
to Mei et al. (2007), the mean distance of 16.5$\pm$0.5 Mpc. Going by the running
median trend, the Bootes filament seems to be slightly curved and galaxy
distances tend to decrease towards the Virgo cluster.

\section{The Hubble flow in the Bootes strip}

For a galaxy with measured distance, its deviation from the unperturbed Hubble 
flow can be characterized by an individual Hubble parameter value $H=(V_{LG})/D$ 
or by a peculiar velocity value $V_{pec}=V_{LG}-H_0\times D$, where $H_0$ is 
fixed to be 72~km$\:$s$^{-1}$~Mpc$^{-1}$.

The upper panel of Figure~3 presents individual $H$ value distribution of the
Bootes strip galaxies lying in the range from 25 up to
100~km$\:$s$^{-1}$~Mpc$^{-1}$. A polyline under the $H$ scale shows the running
median drift with window of $0^h\hspace{-0.4em}.\,5$ along RA. The most common value is
$H=62$~km$\:$s$^{-1}$~Mpc$^{-1}$, which remains almost flat from RA=$13^h\hspace{-0.4em}.\,7$ to
$17^h\hspace{-0.4em}.\,3$. Galaxies in the Virgo infall zone demonstrate clearly a droop of the
$H$ median; to the contrary, in the vicinity of the Local Void $H$ value rises
which is quite expectable since galaxies move away from the void centre
(unfortunately, the galaxy number statistics near the void is poor).

Peculiar velocities distribution of the Bootes strip galaxies is similar (the
lower panel of Figure 3). The main body of the Bootes filament is characterized
by roughly the same value of $V_{pec}\simeq-250$~km$\:$s$^{-1}$ tending to increase
near the Local Void boundary and to decrease significantly (to -600~km$\:$s$^{-1}$)
near the Virgo cluster.

The relation between radial velocities and distances of galaxies in the Bootes
strip is presented in Figure 4, where straight line corresponds to the Hubble
parameter value $H_0=72$~km$\:$s$^{-1}$~Mpc$^{-1}$. Galaxies in the Virgo infall
zone (RA$<14^h\hspace{-0.4em}.\,0$) are plotted as open circles while other single galaxies are
shown as solid ones. Due to distance errors ($\sim$20\%), a nonlinear Malmquist
bias appears in this Hubble diagram: absolute distance errors for farther
galaxies are larger than those for nearer ones, so the whole set of galaxies is
shifted rightwards apparently. To reduce Malmquist bias, we use pairs and groups
of galaxies with distances measured for two or more members. Mean radial
velocities and mean distances for 10 groups and 5 pairs are shown as squares and
triangles, respectively, indicating mean distance error bars. Three groups and a
pair in the Virgo infall zone are plotted with open circles. As can be seen,
in most cases the deviations of groups and pairs from the general Hubble flow are
quite small though exceeding their mean distance errors. The most deviating
group beyond the infall zone is NGC\,5838 with a peculiar velocity of
$\sim500$~km$\:$s$^{-1}$ which is probably due to the influence of the
neighbouring massive group NGC\,5846.

\section{Sub-structures in the Bootes filament}

According to the galaxy grouping criterion (Karachentsev \& Makarov 2008,
Makarov \& Karachentsev 2011) the considered region is populated with 13 groups
and 11 pairs. Their main properties are presented in Tables 2 and 3, respectively.
The columns of Table~2 contain: (1) name of the brightest galaxy in the group;
(2) number of group members with measured radial velocities; (3) mean radial
velocity in the Local Group frame; (4) group distance (in Mpc) corresponding to
the mean distance modulus ($m-M$); (5) dispersion of radial velocities; (6) mean
projected harmonic radius of the group; (7) stellar mass logarithm derived from
the total luminosity of group members in the K- band; (8) projected mass logarithm 
(Heisler et al. 1985):
$$M_p=(32/\pi G)\times(N-3/2)^{-1} \sum^N_{i=1}\Delta V^2_i\times R_i, $$
where $ \Delta V_i$ and $R_i$ are radial velocity and projected distance of the
$i$th galaxy relative to the system centre and $G$ is the gravitational
constant; (9) projected-to-total stellar mass ratio in logarithmic scale; (10)
number of group members with individual distance estimates; (11) the mean-square
difference of distance moduli of the group members; in the case $N_D$ = 1 we 
formally put $\sigma(m-M)=0^m\hspace{-0.4em}.\,4$ which corresponds to distance error of 20\%.

The data on galaxy pairs are given in Table 3 in the same manner. Here,
projected (i.e. orbital) mass is defined as
$$M_p=(16/\pi G)\times \Delta V^2_{12}\times R_{12},$$
where $\Delta V_{12}$ and $R_{12}$ are radial velocity difference and projected
separation between the pair components.

{\em a) Galaxy groups.}

The most prominent structural unit in the Bootes strip is NGC\,5846 group. Being a
compact system densely populated with early types galaxies, NGC\,5846 group is
obviously at the advanced stage of its dynamical evolution. Mulchaey \& Zabludoff
(1998) mention the presence of X-ray emission around NGC\,5846 as the central
galaxy of the group. The population of this group was considered by Zabludoff \&
Mulchaey (1998), Mahdavi et al. (2005), and Eigenthaler \& Zelinger (2010). 
According to Mahdavi et al. (2005), NGC\,5846 group includes about 250 members
with absolute magnitudes brighter than $M_R=-12$ mag while the virial mass of
the group is $8.4\times10^{13}M_{\odot}$. The same authors noticed the group is
significantly isolated along the line of sight: on the 10 sq.\,deg. area
occupied by this group there is not any galaxy lying in the foreground while the
nearest neighbouring galaxies in the background appear only at $V_{LG}\geq
6000$~km$\:$s$^{-1}$. Based on X-ray emission pattern, Mahdavi et al.(2005) 
distinguished two subgroups dominated by two elliptical galaxies: NGC\,5846 and
NGC\,5813. Yet, kinematic characteristics distinguish another group, besides 
NGC\,5846 one. This group includes 9 galaxies dominated by S0 type galaxy
NGC\,5838. The groups around NGC\,5846 and NGC\,5838 have roughly the same distances:
26.4 Mpc and 25.0 Mpc, but differ significantly in their mean radial velocities:
(1803~km$\:$s$^{-1}$ and 1269~km$\:$s$^{-1}$). In such a configuration, the total
dispersion of radial velocities reaching 320~km$\:$s$^{-1}$ (Mahdavi et al. 2005)
reduces to 228~km$\:$s$^{-1}$ (NGC\,5846) and 53~km$\:$s$^{-1}$ (NGC\,5838)
that leads to decrease of the virial mass estimate from
$8.4\times 10^{13}M_{\odot}$ to $4.8\times 10^{13}M_{\odot}$ for the whole
complex of galaxies around NGC\,5846.

Figure 5 represents a close-up view of galaxy distribution in the region of 
NGC\,5846 group. Another populated group around Sb type galaxy NGC\,5746 lies to
the west of NGC\,5846, as well as a group around E type galaxy NGC\,5638, a group
around Sab type galaxy NGC\,5566 and some more poorly populated groups and pairs.
The radii of zero velocity surface for these groups (with virial masses
specified in Table~2) are 2.8~Mpc (NGC\,5846), 1.0~Mpc (NGC\,5838), 2.0~Mpc 
(NGC\,5746) and 1.4~Mpc (NGC\,5566). With weighted average distance $D= 26$~Mpc,
the angular radii of the infall zones are $6.1^{\circ}, 2.2^{\circ},
4.4^{\circ}$ and $3.2^{\circ}$, respectively. Hence, the infall zones around 
the considered groups overlap substantially and the substructures themselves 
can merge eventually into a single dynamical system.

It is worth emphasizing that members of the MK-groups ( Makarov \& Karachentsev,
2011), were found by their radial velocities and projected separations
with regard for galaxy stellar masses ($K$-band luminosities) while their
individual distance estimates were not considered. Since most galaxies in the
Bootes strip have Tully-Fisher distance estimates with an accuracy of $\sim20$\%
or 0.4~mag, the scatter of distance moduli estimates inside a group should be
$\sim0.4$~mag. The data in the last column of Table~2 show that the rms
difference $\sigma (m-M)$ weighted by the number of group members $N_D$ is
0.35~mag. Hence, the applied grouping algorithm does not add a substantial
number of false members into the groups.

{\em b. Binary galaxies.}

The pairs of galaxies from Table~3 are characterized by the median distance of
26~Mpc typical for the whole population of the Bootes strip. The median
radial velocity difference for the components of 11 pairs is only
22~km$\:$s$^{-1}$ giving evidence for their physical connection. The median
projected separation of the components is 180~kpc which is also typical for
dynamically bound pairs. The median projected mass of the binary galaxies is
0.8$\times10^{10}M_{\odot}$ and the typical projected mass-to-total stellar mass
ratio amounts to $M_p/M_*\simeq 7$. For five galaxy pairs having distance
estimates for both components, the $(m-M)$ scatter does not exceed the expected
tolerance of $\sim0.4$ mag.

{\em c. Field galaxies.}

As it was noticed afore in this paper, about 44\% of galaxies in the Bootes
strip are not bound to any system forming the field population. This ratio is
just typical for the whole volume of the Local Supercluster, limited by radial
velocities $V_{LG}<3500$~km$\:$s$^{-1}$ (Makarov \& Karachentsev 2011). Low
luminosity objects and late type galaxies are common among the field population.
This fact fits well into the paradigm in which masses of galaxies and luminosity
of their bulges grow with time due to the repeating hierarchical merging.

As seen from Figure~4, single galaxies in the Bootes strip are more scattered 
in the Hubble diagram than centres of groups and pairs. This is mainly caused by 
distance errors. In spite of the distance scatter, single galaxies on the western 
side of the Bootes strip (open circles) also show the effect of infall towards the 
Virgo cluster similarly to centers of groups and pairs. It should be noted that 
distances of some galaxies measured via Tully-Fisher relation differ significantly 
from those expected from their radial velocities (even taking into account 
the Virgo-centric infall effect). We have eliminated some of these distance 
estimates, for example, for dwarf galaxies SDSS~1430+07 and CGCG~75-063 with 
spurious widths $W_{50}\sim170$~km$\:$s$^{-1}$ attributable to HI flux confusion
from neighbouring bright spirals. However, there are some problematic cases in
Table~1 with a strong discordance between radial velocities and distances. One
of them is an isolated galaxy AGC~238769 with radial velocity of 953~km$\:$s$^{-1}$
and distance estimate of 34.0~Mpc. Another case is a flat galaxy FGC~1642 with
$V_{LG}=1168$~km$\:$s$^{-1}$ and $D=34.6$~Mpc, we have excluded from NGC\,5248
group members when estimating dispersion of $(m-M)$. Significant deviations of 
such galaxies in the Hubble diagram not attributable to their wrong HI
line width or inclination error can give argument for existent large
peculiar motions of these scarce galaxies.

\section{Local density of matter in the Bootes strip}

The virial (projected) mass distribution of galaxy groups (squares) and pairs 
(triangles) in the considered strip versus their total stellar mass is
depicted in Figure~6. There is a positive correlation between virial and 
stellar masses, well-known from other data. While masses are small, the 
significant vertical scatter is caused mainly by projection factors.

According to Jones et al. (2006), the mean density of the stellar matter in the
Universe is $4.6\times 10^8M_{\odot}$ Mpc$^{-3}$ as 
$H_0=72$~km$\:$s$^{-1}$~Mpc$^{-1}$. The global matter density $\Omega_m=0.28$ in
the standard $\Lambda$CDM model with this Hubble parameter is equivalent to
dark-to-stellar matter ratio of $M_{DM}/M_*=97$. This value is shown as a
diagonal in Figure~6. All the groups and all the pairs except one in the Bootes
strip are situated under this line. The sum of virial masses-to-sum of stellar
masses ratio for all the groups and pairs, $\sum M_p/\sum M_*=33$, is plotted
as a cross. The dashed line drawn through the cross indicates the mean mass
density $\Omega_m$ (Bootes) $\simeq 0.09$, which is three times smaller than
the global cosmic density. Here, we did not consider for contribution of single
galaxies, but evidently the field galaxies contribute both to numerator and
denominator of the ratio $\sum M_p/\sum M_*$. Moreover, considering them should 
only reduce slightly this proportion. Thus, the observational data on galaxy 
motions in the Bootes strip give us an argument that this filamentary structure 
does not contain a large amount of dark matter at a level of $\Omega_m\simeq0.28$. 
This statement refers certainly to virial mass estimates based on internal motions 
of galaxies in systems. Yet, supposing that 3--5 times larger mass is hidden in
the Bootes filament between the groups, then the total mass would be about 
$4\times 10^{14}M_{\odot}$, i.e. comparable with the mass of the Virgo cluster, 
and the velocity dispersion for centres of groups and pairs should be 
considerably larger than what is observed.

\section{Concluding remarks}

Reconstruction the 3D cosmic flow field in the Local Universe undertaken by
Courtois et al. (2013) using the Wiener Filter method shows the complex picture
of galaxy motions in the region between the Local Void and the Virgo cluster.
When the spatial location of the Virgo cluster and the pattern of Virgocentric
infall seem to be well-defined, whereas the extent of the Local Void and
the position of its centre stay controversial.

In the diagram presented in the Figure~7 we tried to represent position of the
Bootes filament relative to the Virgo Cluster and Local Void as two main agents
forming the peculiar velocity field in the Bootes strip. We assumed the distance
to the Virgo centre to be 16.5~Mpc (Mei et al. 2007) and the line-of-sight
extent of the Local Void to range from 0 to 20~Mpc with the centre lying halfway
(Nasonova \& Karachentsev 2011). The Local Group with an observer is situated in
the lower left corner while the diagram axes roughly correspond to Supergalactic
axes SGY and SGZ. This diagram illustrates changing proportions between
line-of-sight projected velocities of the Virgocentric infall and of the Void
outflow in different regions of the Bootes filament. This sketch allows to
understand the behaviour of the mean peculiar velocity (Figure~3) along the
filament.

We hope wholesale measurements of galaxy distances both northward and southward
of the Bootes strip will give in the near future more detailed information on
the Local Void geometry and motions of galaxies in its immediate neighborhood.

{\bf Acknowledgments}

\noindent{}This work is supported by the Russian Foundation for Basic Research
(grant no. 13-02-90407) and the State Fund for Fundamental Researches of Ukraine
(grant no. F53.2/15). O.\,G.\,Nasonova thanks the non-profit Dmitry
Zimin’s Dynasty Foundation for the financial support. This research has
made use of NASA/IPAC Extragalactic Database (http://ned.ipac.caltech.edu) which
is operated by the Jet Propulsion Laboratory, California Institute
of Technology, under contract with the National Aeronautics and
Space administration. We acknowledge the usage of the HyperLeda
database (http://leda.univ-lyon1.fr) and SDSS archive (http://www.sdss.org).

\section*{REFERENCES}

{\sloppy

\pagebreak[3]\hangindent=0.5cm\noindent{}Abazajian K.N., Adelman-McCarthy J.K., Agueros M.A., et al. 2009, ApJS, 182, 54

\pagebreak[3]\hangindent=0.5cm\noindent{}Courtois H.M., Pomarede D., Tully R.B., et al. 2013, AJ, 146, 69

\pagebreak[3]\hangindent=0.5cm\noindent{}Dietrich J.P., Werner N., Clowe D., et al., 2012, Nature, 487, 202

\pagebreak[3]\hangindent=0.5cm\noindent{}Eigenthaler P., Zeilinger W.W., 2010, A\&A, 511, 12

\pagebreak[3]\hangindent=0.5cm\noindent{}Haynes M.P., Giovanelli R., Martin A.M., et al. 2011, AJ, 142, 170

\pagebreak[3]\hangindent=0.5cm\noindent{}Heisler J., Tremaine S., Bahcall J.N., 1985, ApJ, 298, 8

\pagebreak[3]\hangindent=0.5cm\noindent{}Jones D.H., Peterson B.A., Colless M., Saunders W., 2006, MNRAS, 369, 25

\pagebreak[3]\hangindent=0.5cm\noindent{}Karachentsev I.D., Tully R.B., E.J.Shaya, et al. 2014, ApJ, 782, 4
\pagebreak[3]\hangindent=0.5cm\noindent{}Karachentsev I.D., Nasonova O.G., 2013, MNRAS, 429, 2677

\pagebreak[3]\hangindent=0.5cm\noindent{}Karachentsev I.D., Nasonova O.G., Courtois H.M., 2013, MNRAS, 429, 2264

\pagebreak[3]\hangindent=0.5cm\noindent{}Karachentsev I.D., 2012, Astrophys. Bull., 67, 115

\pagebreak[3]\hangindent=0.5cm\noindent{}Karachentsev I.D., Nasonova O.G., Courtois H.M., 2011, ApJ, 743, 123

\pagebreak[3]\hangindent=0.5cm\noindent{}Karachentsev I.D., Nasonova O.G., 2010, MNRAS, 405, 1075

\pagebreak[3]\hangindent=0.5cm\noindent{}Karachentsev I.D., Kashibadze O.G., Makarov D.I., Tully R.B., 2009, MNRAS, 393, 1265

\pagebreak[3]\hangindent=0.5cm\noindent{}Karachentsev I.D., Makarov D.I., 2008, Astrophys. Bulletin, 63, 299

\pagebreak[3]\hangindent=0.5cm\noindent{}Karachentsev I.D., Tully R.B., Dolphin A., et al. 2007, AJ, 133, 504

\pagebreak[3]\hangindent=0.5cm\noindent{}Karachentsev I.D., Kashibadze O.G., 2006, Ap, 49, 3

\pagebreak[3]\hangindent=0.5cm\noindent{}Mahdavi A., Trentham N., Tully R.B., 2005, AJ, 130, 1502

\pagebreak[3]\hangindent=0.5cm\noindent{}Makarov D.I., Karachentsev I.D., 2011, MNRAS, 412, 2498

\pagebreak[3]\hangindent=0.5cm\noindent{}Mei S., Blakeslee J.P., Cote P., et al. 2007, ApJ, 655, 144

\pagebreak[3]\hangindent=0.5cm\noindent{}Mulchaey J.S., Zabludoff A.I., 1998, ApJ, 496, 73

\pagebreak[3]\hangindent=0.5cm\noindent{}Nasonova O.G., Karachentsev I.D., 2011, Ap, 54, 1

\pagebreak[3]\hangindent=0.5cm\noindent{}Nasonova O.G., de Freitas Pacheco J.A., Karachentsev I.D., 2011, A\&A, 532A, 104

\pagebreak[3]\hangindent=0.5cm\noindent{}Roychowdhury S., Chengalur J.N., Karachentsev I.D., Kaisina E. I., 2013, MNRAS, 436L, 104

\pagebreak[3]\hangindent=0.5cm\noindent{}Sorce J.G., Courtois H.M., Gottloeber S. et al. 2014, MNRAS, 437, 3586

\pagebreak[3]\hangindent=0.5cm\noindent{}Spergel D.N. et al. 2007, ApJS, 170, 377

\pagebreak[3]\hangindent=0.5cm\noindent{}Tonry J.L., Dressler A., Blakeslee J.P., et al., 2001, ApJ, 546, 681

\pagebreak[3]\hangindent=0.5cm\noindent{}Tully R.B., Rizzi L., Shaya E.J., et al. 2009, AJ, 138, 323

\pagebreak[3]\hangindent=0.5cm\noindent{}Tully R.B., 1987, ApJ, 321, 280

\pagebreak[3]\hangindent=0.5cm\noindent{}Tully R.B., Fisher R.J., 1977, A\&A, 54, 661

\pagebreak[3]\hangindent=0.5cm\noindent{}Vennik J., 1984, Tartu Astron. Obs. Publ., 73, 1

\pagebreak[3]\hangindent=0.5cm\noindent{}Whiting A.B., 2006, AJ, 131, 1996

\pagebreak[3]\hangindent=0.5cm\noindent{}Zabludoff A.I., Mulchaey J.S., 1998, ApJ, 496, 39

\pagebreak[3]\hangindent=0.5cm\noindent{}Zwaan M.A., Staveley-Smith L., Koribalski B.S. et al., 2003, AJ, 125, 2842

}

\clearpage
\onecolumn
\begin{figure}
\includegraphics[height=1.0\textwidth,keepaspectratio,angle=270]{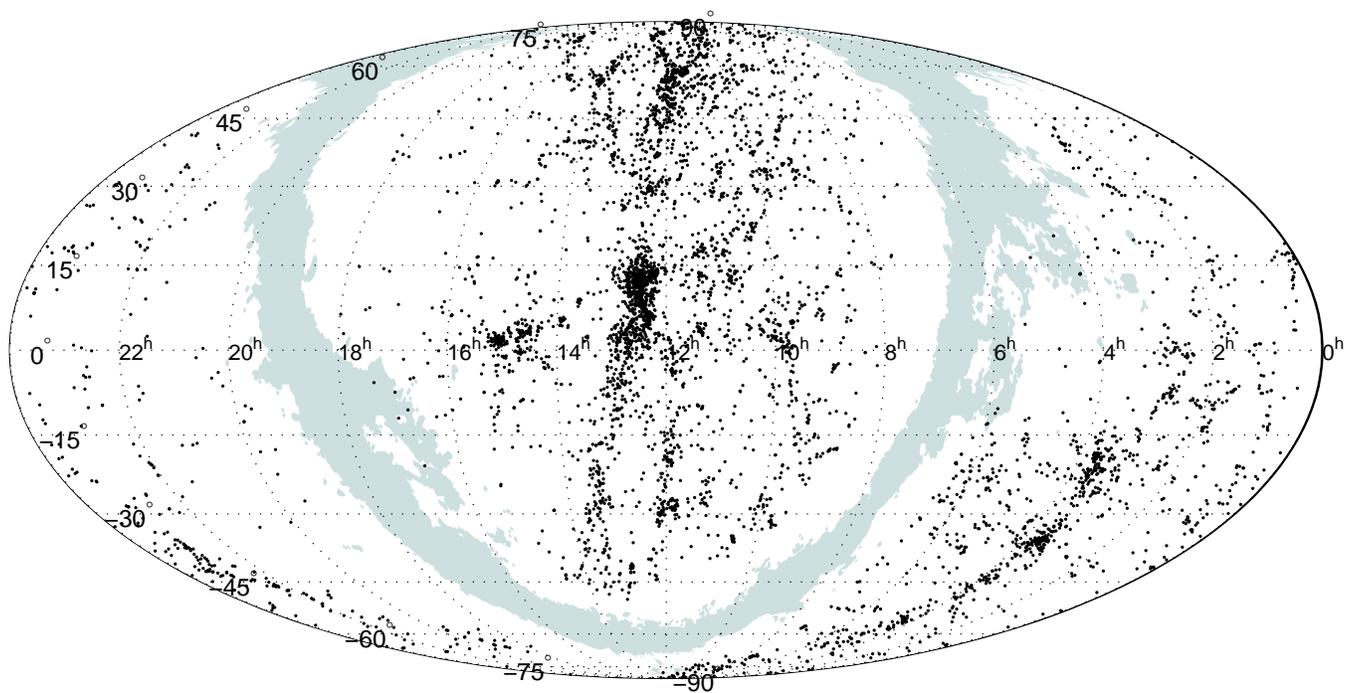}
\caption{Sky distribution of Local Supercluster galaxies in equatorial coordinates. 
The region of strong absorption along the Milky Way is filled with gray.}
\end{figure}

\begin{figure}
\includegraphics[height=1.0\textwidth,keepaspectratio,angle=270]{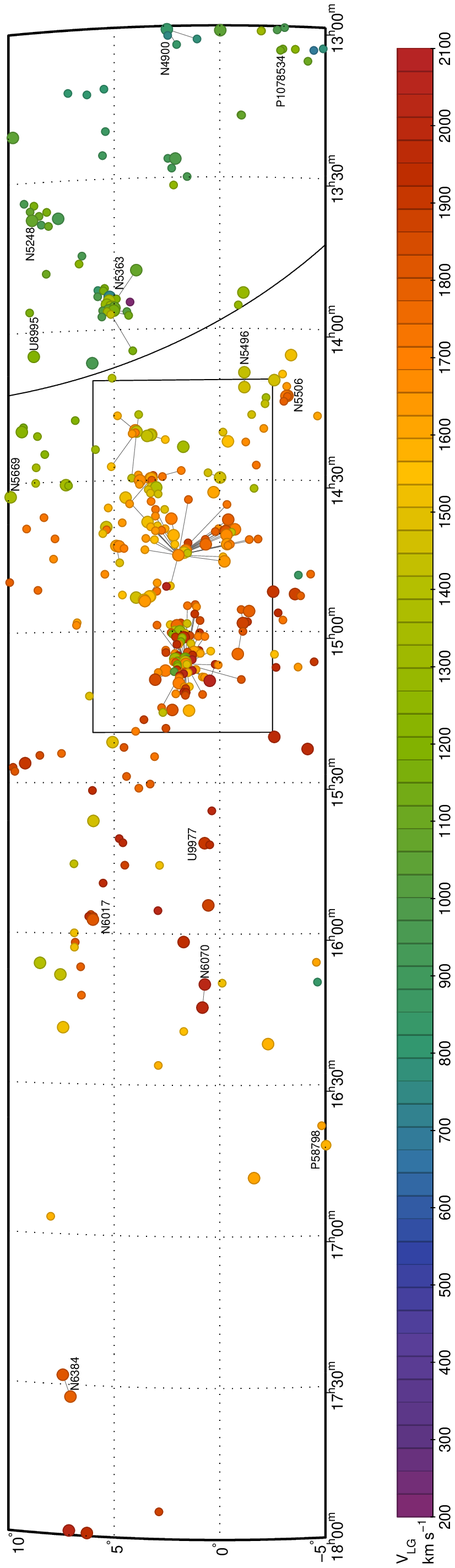}
\vspace{1.2cm}\par
\includegraphics[height=1.0\textwidth,keepaspectratio,angle=270]{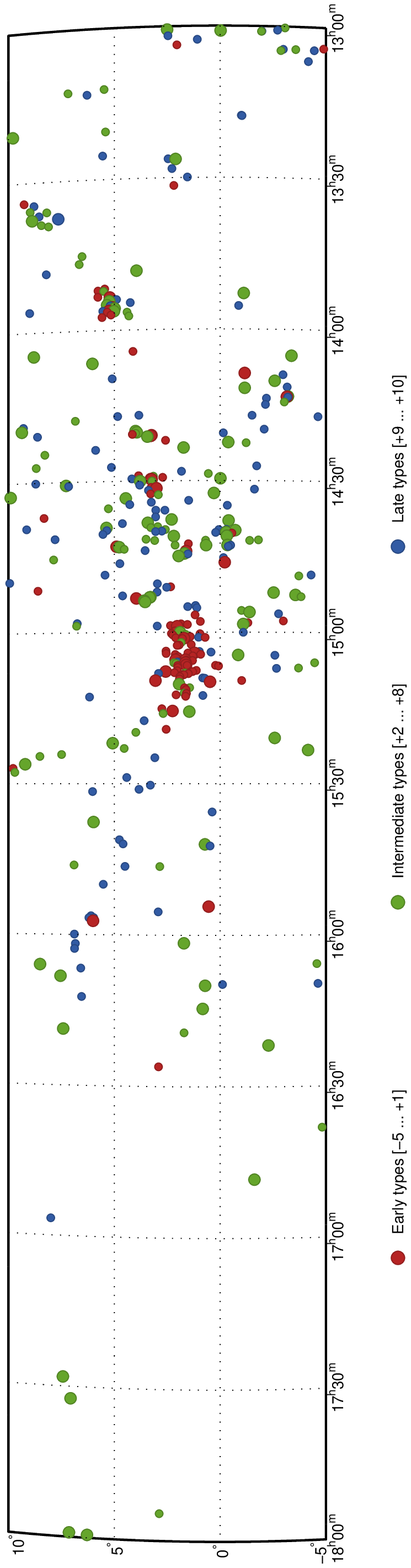}\par
\vspace{0.9cm}\par
\includegraphics[height=1.0\textwidth,keepaspectratio,angle=270]{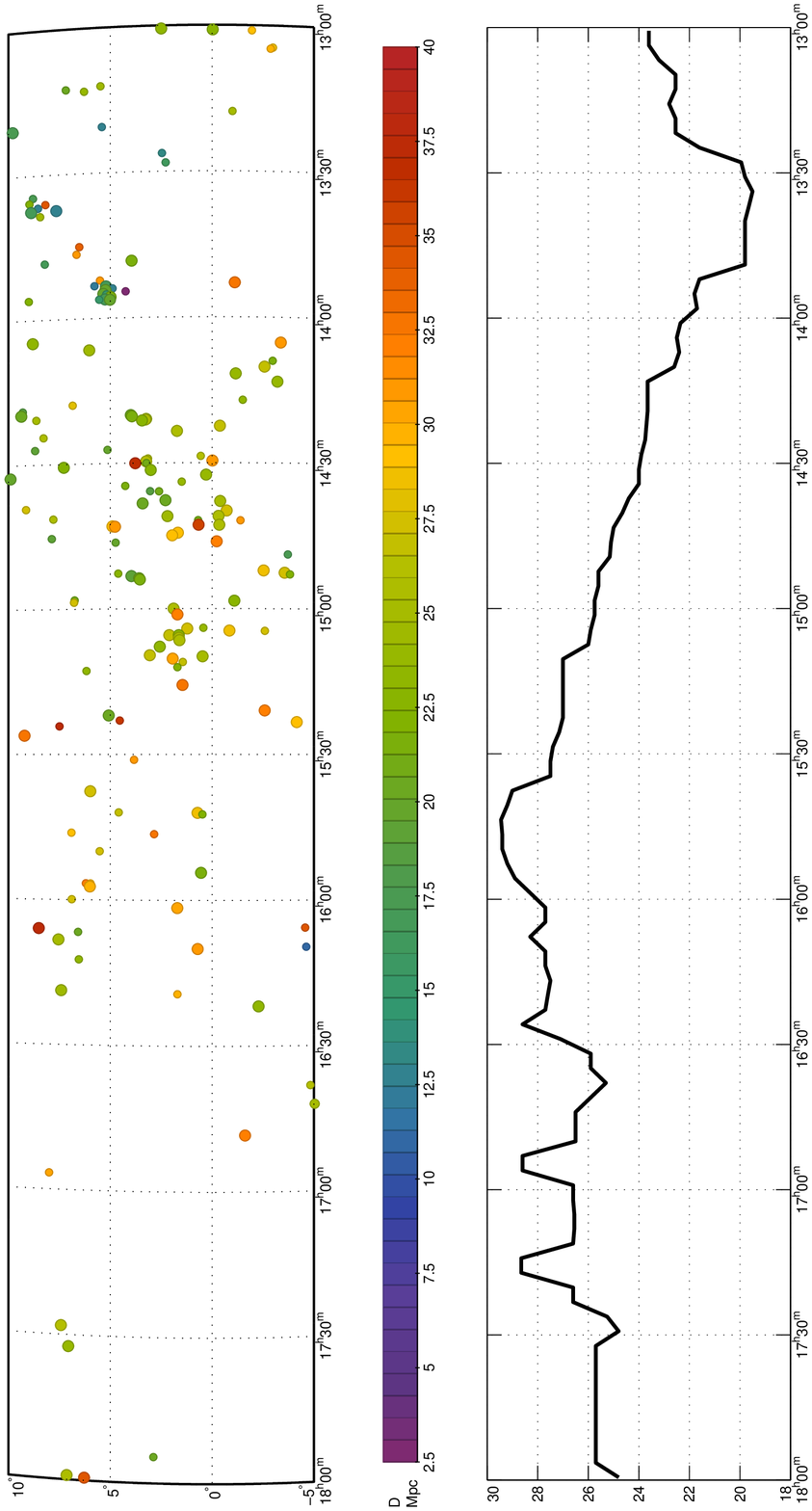}\par
\vspace{1.0cm}\par
\caption{Radial velocity (upper panel), morphological type (middle panel) and
distance (lower panel) distributions of galaxies in the Bootes strip. Dwarf
galaxies with absolute magnitudes $M_B>-17.0$ are plotted with small circles.
A~polyline under the lower panel represents the running median of distance with
window of $0^h\hspace{-0.4em}.\,5$.}
\end{figure}

\begin{figure}
\includegraphics[height=1.0\textwidth,keepaspectratio,angle=270]{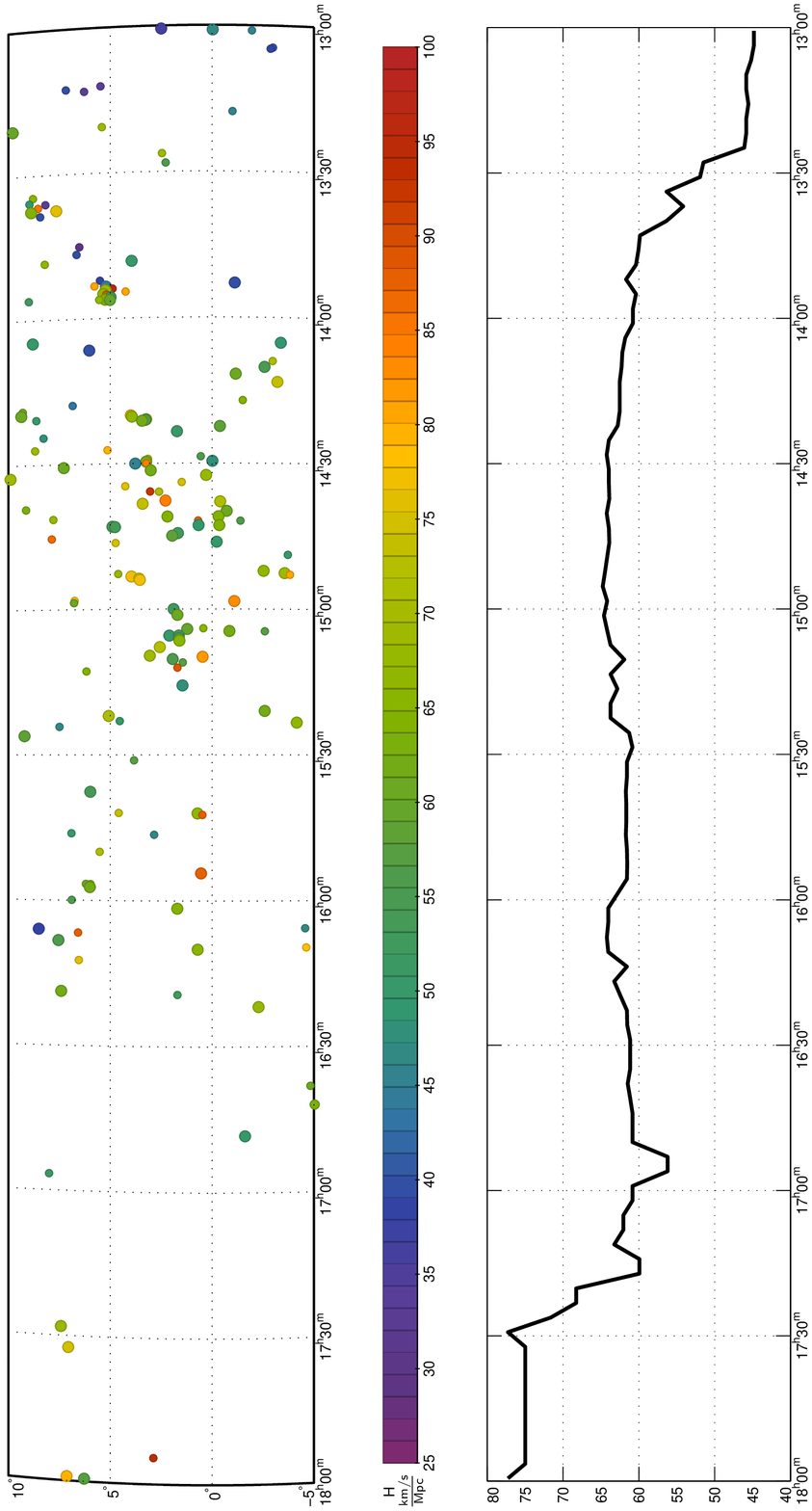}\par
\vspace{1.2cm}\par
\includegraphics[height=1.0\textwidth,keepaspectratio,angle=270]{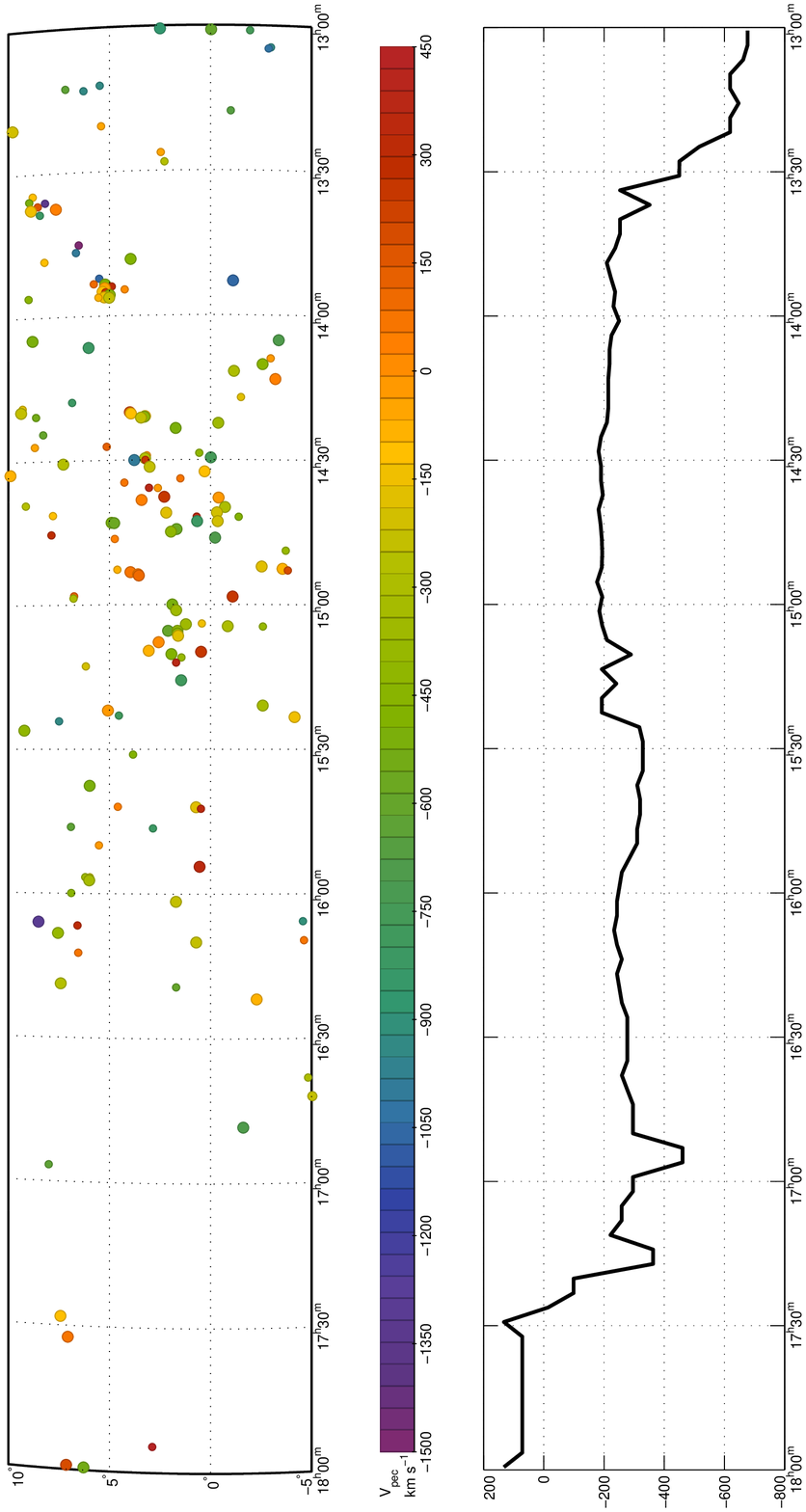}\par
\vspace{1.0cm}\par
\caption{Individual $H$ value (upper panel) and peculiar velocity distributions
of galaxies in the Bootes strip. Polygonal lines under the panels correspond to
running medians with window of $0^h\hspace{-0.4em}.\,5$. Dwarf galaxies are plotted with small
circles.}
\end{figure}

\begin{figure}
\includegraphics[height=1.0\textwidth,keepaspectratio,angle=270]{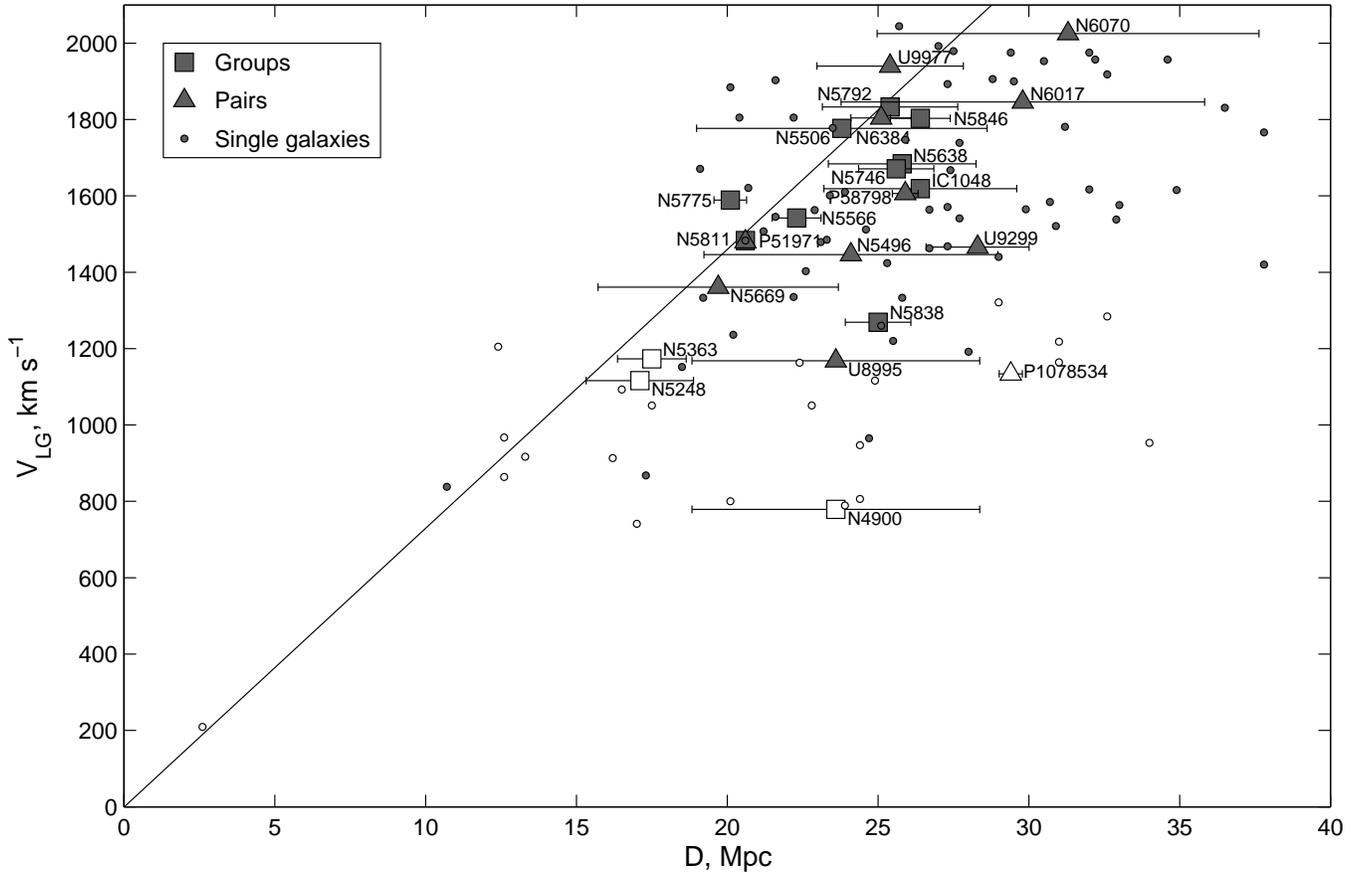}
\caption{Hubble digram for group centres (squares), pair centres (triangles) and
single galaxies (circles). Galaxies and systems in the Virgo infall zone with
RA$<14^h\hspace{-0.4em}.\,0$ are plotted with open symbols.}
\end{figure}

\begin{figure}
\includegraphics[height=1.0\textwidth,keepaspectratio,angle=270]{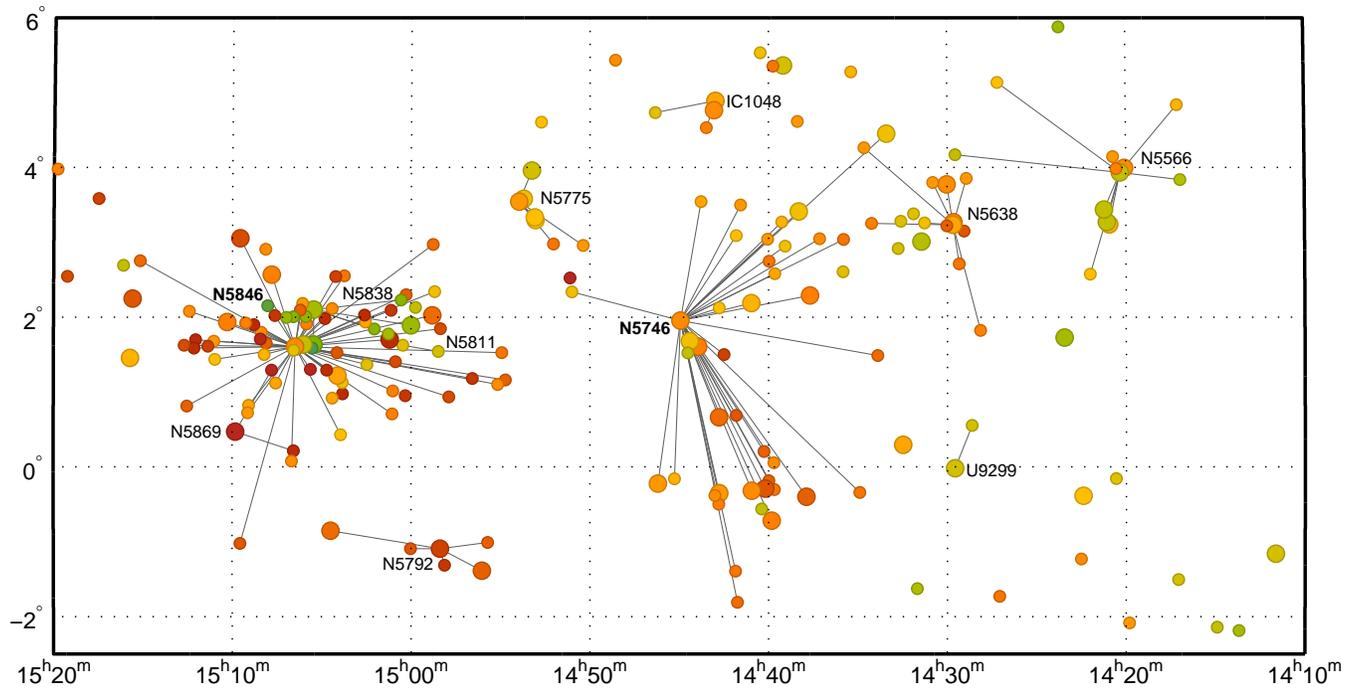}
\caption{Close-up view of the Bootes strip region containing NGC\,5846 and
NGC\,5746 groups. Members of groups and pairs are linked with their dominating
galaxies.}
\end{figure}

\begin{figure}
\includegraphics[height=0.8\textwidth,keepaspectratio,angle=270]{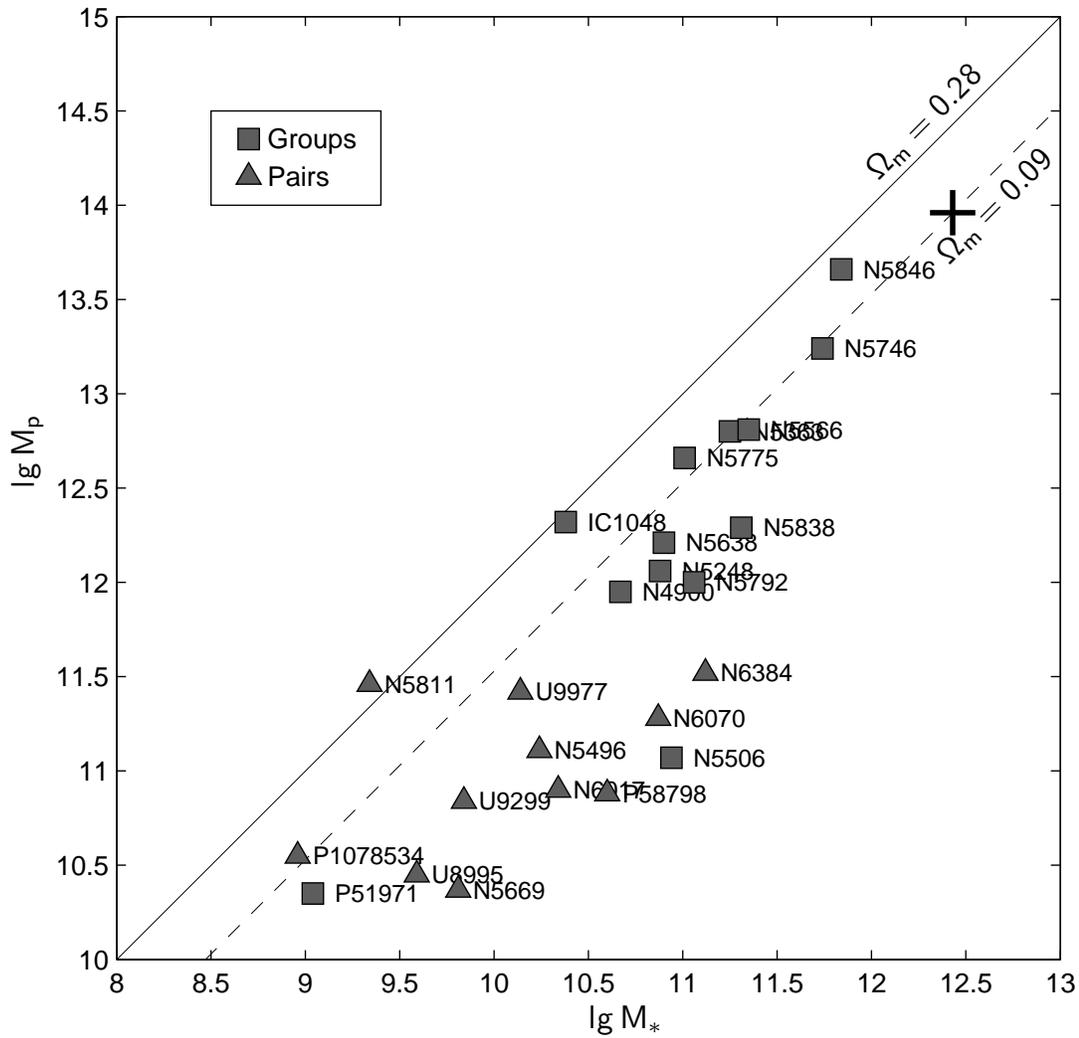}
\caption{The relation between projected (virial) mass and total stellar mass for
galaxy groups (squares) and pairs (triangles). A cross marks virial and stellar
mass values for all the Boots strip. Solid and dashed lines indicate the global
and the local mean matter density.}
\end{figure}

\begin{figure}
\includegraphics[height=0.8\textwidth,keepaspectratio,angle=270]{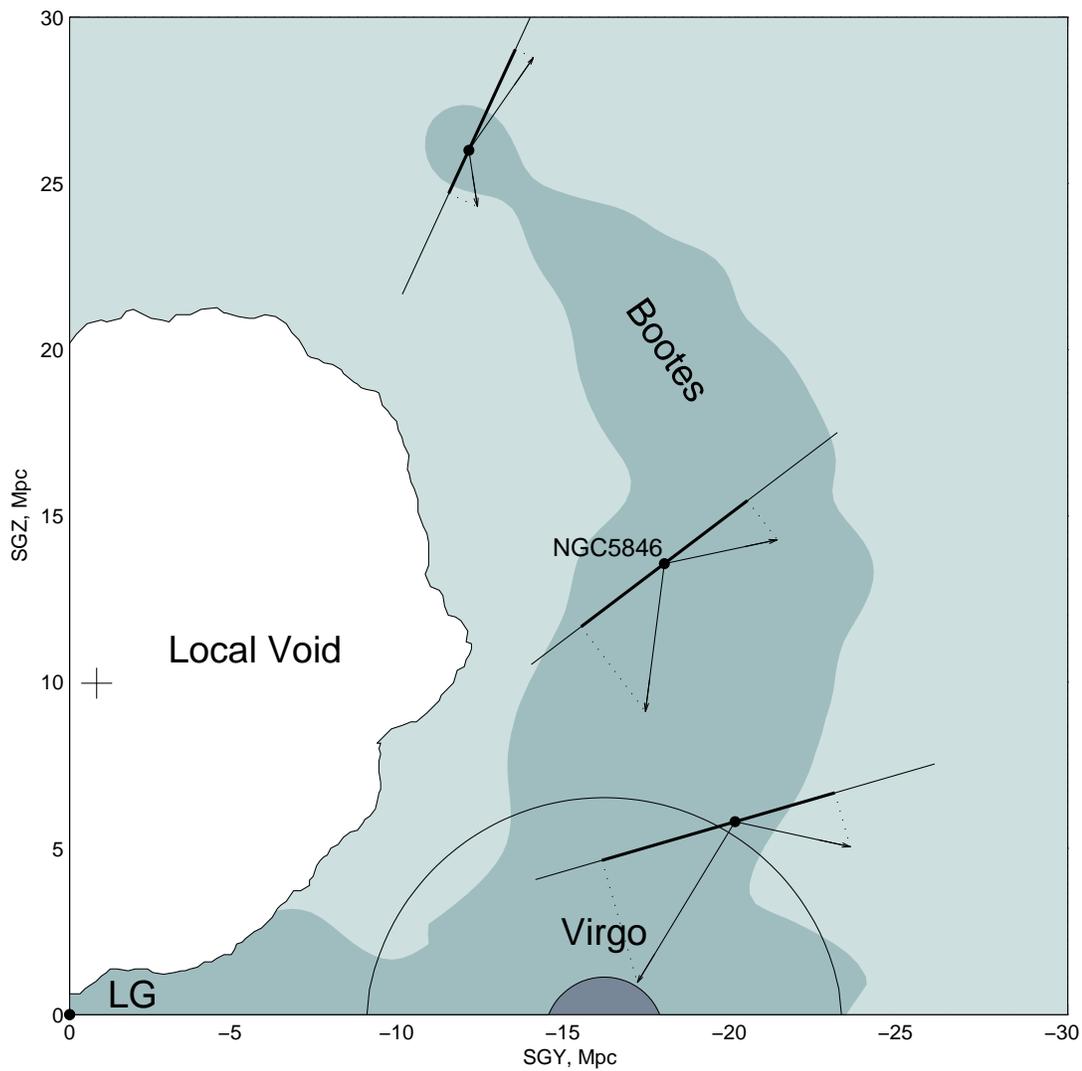}
\caption{The diagram illustrating the position of the Bootes filament relative
to the Virgo cluster and the Local Void. The inner and the outer radii around the
Virgo cluster correspond to the virial zone and the zero velocity surface radius
$R_0$. The projected velocities of the Virgocentric infall and of the Void
outflow have different values in different regions of the Bootes filament. An
observer (LG) is situated in the lower left corner of the diagram.}
\end{figure}

\clearpage
\begin{center}\small{}
\begin{longtable}{lcccrrll}
\caption{Galaxies with $V_{LG} < 2000$~km~s$^{-1}$ in the Bootes strip: RA =
[13.0, 18.0]$^h$, Dec = [$-$5, +10]$^{\circ}$}\\
\hhline{========}
Object &  RA (2000.0) DEC  & $B_t$&    $D_{Mpc}$& $ V_{LG}$& err&  $T$&     Group \\
\hhline{========}\endfirsthead
\hhline{========}
Object &  RA (2000.0) DEC  & $B_t$&    $D_{Mpc}$& $ V_{LG}$& err&  $T$&     Group \\
\hhline{========}\endhead
\hhline{--------}\endfoot
\hhline{========}\endlastfoot
PGC044733 & J130017.5-030359& 15.88 &         &  963 & 21 & Sm &          \\
NGC4900   & J130039.2+023004& 11.91 & 23.6 sn &  832 &  8 & Sc &   N4900  \\
PGC1084547& J130045.7-024304& 18.12 &         &  991 & 89 & Ir &          \\
NGC4904   & J130058.7-000139& 12.67 & 22.8 tf & 1051 &  8 & Sc &          \\
PGC214054 & J130100.8-015834& 17.39 &         & 1216 & 15 & Im &          \\
UGC08127  & J130103.7-015712& 15.57 & 29.0 TF & 1321 & 10 & Sm &          \\
PGC1227695& J130153.4+022738& 17.96 &         &  746 &  8 & BCD&   N4900  \\
PGC045019 & J130240.8+010427& 17.47 &         &  787 & 20 & BCD&   N4900  \\
PGC135818 & J130344.2+020224& 17.18 &         &  844 & 49 & dS0&   N4900  \\
PGC170228 & J130412.1-045327& 14.83 &         &  765 & 64 & S0 &          \\
UGCA322   & J130431.2-033421& 14.34 &         & 1211 &  9 & Sm &          \\
PGC1078534& J130431.9-025917& 16.28 & 28.8 TF & 1148 & 35 & Im &   P1078534\\
PGC3271456& J130436.5-042706& 18.93 &         &  683 & 90 & Ir &           \\
PGC1080976& J130446.5-025216& 16.37 & 29.8 TF & 1113 & 89 & Scd&   P1078534\\
PGC3271328& J130646.8-041021& 16.43 &         & 1109 & 64 & BCD&           \\
UGC08276  & J131206.4+052832& 17.12 & 24.4 TF &  806 &  7 & Sm &           \\
UGC08285  & J131233.3+071103& 15.0  & 20.1 TF &  800 &  9 & Sm &           \\
AGC238737 & J131304.4+061707& 18.3  & 23.9 TF &  789 & 21 & Im &           \\
PGC046306 & J131742.5-010006& 15.97 & 24.9 TF & 1116 & 13 & Im &           \\
PGC1128365& J131746.2-010215& 16.56 &         & 1070 & 64 & Im &           \\
UGC08382  & J132032.1+052428& 15.22 & 12.6 TF &  864 &  7 & Sm &           \\
UGC08385  & J132038.1+094714& 14.08 & 17.5 TF & 1051 &  5 & Sm &           \\
AGC238691 & J132517.6+053236& 17.7  &         &  893 & 15 & Ir &           \\
PGC135826 & J132615.9+022731& 16.80 & 13.3 TF &  917 &  7 & Im &           \\
NGC5147   & J132619.6+020603& 12.27 &         &  978 &  8 & Sm &           \\
KKSG 64   & J132812.2+021643& 17.44 & 16.2 TF &  913 &  7 & Im &           \\
PGC135828 & J132955.7+013239& 16.55 &         &  939 & 21 & Im &           \\
PGC135829 & J133133.9+021115& 17.79 &         & 1243 & 89 & dE &           \\
SDSSJ13340& J133406.9+091543& 17.70 &         &  958 & 39 & dE &           \\
KKH 84    & J133437.9+084737& 15.78 & 17.6 TF & 1156 &  5 & Im &   N5248   \\
UGC08575  & J133545.6+085809& 15.5  & 22.3 TF & 1089 &  4 & Sd &   N5248   \\
FGC1642   & J133602.4+081108& 16.8  & 34.8 TF & 1168 &  5 & Sd &   N5248   \\
CGCG73-051& J133643.7+083248& 15.88 & 12.7 TF & 1091 &  5 & Ir &   N5248   \\
UGC08614  & J133726.2+073842& 13.21 & 12.6 TF &  967 &  7 & Im &           \\
NGC5248   & J133732.1+085306& 10.88 & 16.9 tf & 1078 &  4 & Sbc&   N5248   \\
UGC08629  & J133830.6+082632& 15.71 & 24.4 TF &  947 &  4 & Sm &           \\
AGC233601 & J133850.8+080629& 17.5  &         & 1113 &  5 & Sm &           \\
AGC238769 & J134506.5+063110& 16.9  & 34.0 TF &  953 &  5 & Sdm&           \\
AGC238771 & J134641.0+063915& 17.11 & 31.0 TF & 1218 & 47 & Sm &           \\
NGC5300   & J134816.0+035703& 13.06 & 21.6 tf & 1086 &  8 & Sc &   N5363   \\
AGC713655 & J134822.8+081241& 17.8  & 16.5 TF & 1093 & 20 & BCD&           \\
PGC1283560& J135143.0+052647& 16.20 &         & 1165 & 24 & dE &   N5363   \\
SDSSJ13520& J135205.7+054554& 17.11 &         &  825 &  5 & dS0&           \\
AGC232162 & J135210.8+053013& 17.5  & 31.0 TF & 1164 & 33 & Sm &           \\
NGC5334   & J135254.5-010652& 12.97 & 32.6 tf & 1284 &  7 & Sc &           \\
UGC08799  & J135319.8+054618& 16.32 & 12.1 sbf&  974 & 28 & dE &   N5363   \\
NGC5338   & J135326.5+051228& 13.99 & 17.0 TF &  741 & 10 & S0 &           \\
AGC238709 & J135353.8+045316& 18.0  & 12.4 TF & 1205 & 14 & Ir &           \\
NGC5348   & J135411.2+051338& 14.18 & 19.8 tf & 1376 &  7 & Sc &   N5363   \\
KKH 86    & J135433.5+041440& 17.08 &  2.6 rgb&  209 &  5 & Ir &           \\
NGC5356   & J135458.4+052001& 13.63 & 19.5 tf & 1300 &  7 & Sb &   N5363   \\
PGC1277985& J135502.7+050525& 17.01 &         & 1322 &  7 & dEn&   N5363   \\
AGC232141 & J135504.5+051122& 17.18 & 14.8 TF & 1329 &  5 & BCD&   N5363   \\
PGC3303692& J135522.4-005237& 18.78 &         & 1289 & 89 & BCD&           \\
NGC5360   & J135538.7+045906& 14.5  & 21.5 TF & 1097 & 10 & Sm &   N5363   \\
F721-v2   & J135558.0+085947& 18.0  & 22.4 TF & 1163 & 23 & Ir &           \\
NGC5363   & J135607.3+051517& 11.10 & 16.6 TF & 1066 &  7 & S0 &   N5363   \\
AGC232142 & J135609.4+053234& 17.38 & 15.1 TF & 1011 &  8 & Ir &   N5363   \\
NGC5364   & J135612.0+050052& 11.19 & 19.5 tf & 1167 &  6 & Sbc&   N5363   \\
SDSSJ13562& J135621.3+051944& 17.37 &         & 1322 & 68 & dE &   N5363   \\
UGC08857  & J135626.6+042348& 15.26 &         & 1014 & 54 & Sab&   N5363   \\
PGC049602 & J135655.6+050907& 15.82 &         & 1444 & 25 & dEn&   N5363   \\
PGC1266441& J135714.1+041826& 17.1  &         & 1124 & 30 & Sm &   N5363   \\
PGC1285591& J135723.6+053427& 16.3  &         &  984 & 36 & Sph&   N5363   \\
UGC08986  & J140415.9+040644& 15.03 &         & 1162 & 16 & dEn&   N5363   \\
PGC2807027& J140441.6+084716& 18.0  &         & 1145 &  3 & Ir &   U8995   \\
UGC08995  & J140447.3+084803& 14.83 & 23.6 tf & 1190 &  7 & Scd&   U8995   \\
PGC050229 & J140511.1-032211& 13.86 & 30.9 TF & 1521 &  7 & Sa &           \\
NGC5470   & J140631.7+060144& 14.12 & 24.7 tf &  965 & 10 & Sb &           \\
IC0976    & J140843.3-010942& 14.05 &         & 1436 & 16 & S0 &   N5496   \\
IC0978    & J140858.1-025826& 15.18 & 21.6 TF & 1545 & 12 & BCD&           \\
AGC242151 & J140937.6+050546& 16.3  &         & 1396 & 25 & Ir &           \\
UGC09057  & J141012.9-023433& 14.34 & 26.7 TF & 1463 &  8 & Sd &           \\
PGC1075297& J141124.0-031003& 18.08 &         & 1644 & 64 & BCD&           \\
NGC5496   & J141138.1-010938& 13.52 & 24.1 tf & 1455 &  6 & Scd&   N5496   \\
NGC5506   & J141314.9-031227& 12.81 & 23.8 tf & 1760 & 10 & Sa &   N5506   \\
NGC5507   & J141319.9-030856& 13.43 &         & 1751 & 18 & S0 &   N5506   \\
PGC1074769& J141324.1-031156& 18.35 &         & 1786 &  6 & BCD&   N5506   \\
KDG 230   & J141341.0-021112& 17.50 &         & 1357 &  7 & Ir &           \\
KKs 65    & J141422.2-030156& 15.85 &         & 1811 &  6 & Sm &   N5506   \\
PGC1100661& J141454.2-020823& 17.98 &         & 1448 &  6 & BCD&           \\
KKR 5     & J141657.3+035008& 16.91 &         & 1409 &  5 & Im &   N5566   \\
KKR 6     & J141704.2-013020& 17.48 & 23.1 TF & 1479 &  6 & Im &           \\
AGC242152 & J141707.5+045013& 18.13 &         & 1582 &  4 & BCD&   N5566   \\
PGC1055685& J141710.4-043715& 15.69 &         & 1623 & 89 & BCD&           \\
AGC243881 & J141750.7+065023& 17.0  & 28.0 TF & 1200 & 38 & Sm &           \\
AGC244129 & J141853.5+091729& 17.9  & 18.5 TF & 1152 &  7 & BCD&           \\
UGC09169  & J141944.6+092144& 14.10 & 20.2 TF & 1236 &  4 & Sm &           \\
PGC3304682& J141948.6-020455& 17.23 &         & 1650 & 14 & Im &           \\
NGC5560   & J142004.9+035931& 13.18 & 20.5 tf & 1671 &  5 & Sb &   N5566   \\
NGC5566   & J142019.9+035600& 11.42 & 21.8 tf & 1449 &  5 & Sab&   N5566   \\
NGC5569   & J142032.1+035859& 15.12 &         & 1723 &  5 & Scd&   N5566   \\
PGC1150546& J142033.9-000918& 18.75 &         & 1498 & 89 & Ir &           \\
PGC3125420& J142043.5+040837& 17.24 &         & 1648 & 44 & dE &   N5566   \\
AGC714055 & J142044.5+083736& 16.66 & 25.1 TF & 1260 & 20 & Ir &           \\
NGC5574   & J142056.0+031417& 13.29 & 23.9 sbf& 1599 & 10 & S0 &   N5566   \\
NGC5576   & J142103.7+031616& 11.79 & 25.5 sbf& 1427 & 17 & E  &   N5566   \\
NGC5577   & J142113.1+032609& 13.54 & 23.0 tf & 1430 &  5 & Sbc&   N5566   \\
PGC1231137& J142200.1+023422& 18.19 &         & 1568 & 27 & dE &   N5566   \\
NGC5584   & J142223.8-002315& 12.73 & 26.7 sn & 1564 &  6 & Scd&           \\
PGC1123741& J142230.7-011346& 17.20 &         & 1673 & 54 & Sm &           \\
UGC09215  & J142327.2+014334& 13.75 & 25.8 tf & 1333 &  4 & Sd &           \\
AGC244252 & J142343.7+055237& 17.33 &         & 1325 &  7 & BCD&           \\
UGC09225  & J142424.3+081634& 15.4  & 25.5 TF & 1220 &  4 & Sm &           \\
UGC09249  & J142659.8+084101& 15.01 & 19.2 tf & 1333 &  3 & Sdm&           \\
PGC1111622& J142704.8-014347& 18.09 &         & 1745 & 37 & BCD&           \\
UGC09252  & J142710.8+050801& 15.67 & 19.6 TF & 1553 &  6 & Ir &   N5566   \\
PGC1206999& J142808.6+014926& 18.20 &         & 1706 & 70 & Ir &   N5638   \\
PGC051719 & J142837.8+003311& 15.41 & 26.1 TF & 1459 & 19 & Scd&   U9299   \\
SDSSJ14285& J142855.9+035106& 18.17 &         & 1638 & 32 & dEn&   N5638   \\
UGC09285  & J142903.9+030856& 15.41 & 25.6 tf & 1820 & 22 & Sc &   N5638   \\
PGC1235336& J142919.6+024238& 17.18 &         & 1738 &  5 & dE &   N5638   \\
SDSSJ14293& J142933.5+041014& 18.29 &         & 1448 & 24 & Ir &   N5566   \\
UGC09299  & J142934.6-000105& 14.57 & 30.8 tf & 1473 &  5 & Sd &   U9299   \\
NGC5636   & J142939.0+031559& 13.85 &         & 1693 & 10 & S0 &   N5638   \\
NGC5638   & J142940.4+031400& 12.16 & 26.4 sbf& 1624 &  5 & E  &   N5638   \\
UGC09310  & J143001.1+031314& 15.13 & 21.2 TF & 1794 &  5 & Sm &   N5638   \\
IC1022    & J143001.8+034622& 15.20 & 37.3 tf & 1672 &  9 & Sbc&   N5638   \\
SDSSJ14300& J143007.2+084216& 18.56 &         & 1395 & 36 & Ir &           \\
NGC5645   & J143039.4+071630& 12.90 & 22.2 tf & 1335 &  3 & Sd &           \\
SDSSJ14304& J143048.2+034750& 17.78 &         & 1637 & 24 & Ir &   N5638   \\
SDSSJ1430 & J143048.7+070926& 18.09 &         & 1288 & 68 & Ir &           \\
PGC1248433& J143116.5+031524& 17.04 &         & 1536 & 27 & dEn&   N5638   \\
IC1024    & J143127.1+030030& 14.00 & 22.6 TF & 1403 &  7 & S0 &           \\
PGC1114278& J143141.2-013729& 17.99 &         & 1381 & 20 & Ir &           \\
SDSSJ14315& J143153.0+032249& 18.53 &         & 1491 & 16 & Im &   P051971 \\
CGCG75-063& J143220.8+095600& 15.57 &         & 1373 &  2 & BCD&   N5669   \\
UGC9348   & J143228.5+001739& 14.4  & 23.9 TF & 1610 & 10 & Sb &           \\
PGC1248951& J143235.5+031651& 16.80 &         & 1490 & 29 & dS0&   P051971 \\
NGC5669   & J143243.8+095329& 12.97 & 19.7 tf & 1349 &  5 & Scd&   N5669   \\
PGC051971 & J143245.1+025453& 15.25 &         & 1469 &  5 & Sd &   P051971 \\
NGC5668   & J143324.3+042701& 12.25 &         & 1533 &  3 & Scd&   N5746   \\
KKSG 66   & J143353.5+012912& 17.40 & 23.7 TF & 1776 &  5 & Ir &   N5746   \\
SDSSJ143414&J143414.3+031502& 17.94 &         & 1697 & 58 & Ir &   N5638   \\
UGC09380  & J143439.1+041542& 15.15 & 21.7 tf & 1649 &  7 & Im &   N5638   \\
PGC1145728& J143454.2-002033& 17.29 &         & 1734 &  4 & BCD&   N5746   \\
UGC09385  & J143522.8+051636& 16.4  &         & 1599 &  5 & Sm &           \\
UGC09392  & J143548.5+030218& 17.5  & 18.6 TF & 1734 &  6 & Ir &   N5746   \\
PGC135851 & J143550.1+023619& 16.80 & 21.2 TF & 1507 & 20 & Ir &           \\
PGC1344367& J143701.3+081905& 17.22 &         & 1743 & 43 & En &           \\
PGC052256 & J143708.9+030250& 16.91 &         & 1663 & 39 & Im &   N5746   \\
NGC5690   & J143741.1+021727& 12.48 & 20.2 tf & 1704 &  5 & Sc &   N5746   \\
NGC5691   & J143753.3-002356& 12.90 & 25.2 TF & 1810 &  9 & Sa &   N5746   \\
NGC5692   & J143818.1+032437& 13.46 & 20.5 TF & 1517 & 24 & Sbc&   N5746   \\
SDSSJ14382& J143822.6+043648& 17.93 &         & 1659 &  4 & Ir &           \\
UGC09432  & J143904.1+025653& 16.06 &         & 1529 &  7 & Sm &   N5746   \\
NGC5701   & J143911.1+052148& 12.08 &         & 1470 &  2 & Sa &           \\
AGC714204 & J143912.4+090806& 16.64 & 27.7 TF & 1739 & 53 & Im &           \\
PGC1248765& J143915.5+031621& 16.95 &         & 1597 & 37 & Sm &   N5746   \\
PGC1231422& J143939.3+023455& 17.39 &         & 1603 & 37 & Sm &   N5746   \\
PGC135857 & J143940.7-001809& 17.31 &         & 1749 & 15 & Ir &   N5746   \\
PGC1155970& J143942.7+000322& 18.21 &         & 1685 & 22 & Ir &   N5746   \\
AGC241893 & J143944.5+052112& 17.0  &         & 1727 & 14 & Ir &           \\
NGC5705   & J143949.8-004306& 14.19 & 27.7 tf & 1698 &  5 & Sd &   N5746   \\
PGC1236341& J143958.3+024453& 17.83 &         & 1727 & 22 & Ir &           \\
PGC135858 & J143959.9-001110& 18.14 &         & 1802 &  3 & Ir &   N5746   \\
SDSSJ14400& J144002.9+030228& 18.54 &         & 1635 &  6 & BCD&   N5746   \\
NGC5713   & J144011.5-001721& 12.18 &         & 1841 &  8 & Sbc&   N5746   \\
PGC1159795& J144015.3+001224& 16.93 &         & 1806 &  4 & Ir &   N5746   \\
PGC1140314& J144023.3-003344& 18.24 &         & 1438 & 16 & dE &           \\
SDSSJ14402& J144027.6+053155& 18.09 &         & 1526 & 30 & BCD&           \\
NGC5719   & J144056.4-001905& 13.27 & 26.0 tf & 1676 &  5 & Sa &   N5746   \\
NGC5725   & J144058.4+021113& 14.56 & 24.4 TF & 1580 &  7 & Scd&   N5746   \\
AGC249303 & J144119.2+074735& 16.1  & 25.9 TF & 1747 &  2 & BCD&           \\
PGC1253386& J144133.7+032948& 17.41 &         & 1634 &  4 & Sm &   N5746   \\
UGC09469  & J144144.3-014827& 15.59 &         & 1774 &  7 & Sm &   N5746   \\
SDSSJ14414& J144148.7+030523& 16.34 &         & 1549 & 47 & Sm &   N5746   \\
UGC09470  & J144148.7+004113& 15.12 & 20.7 TF & 1834 &  6 & Sd &   N5746   \\
UGC09472  & J144151.0-012333& 15.84 & 30.9 TF & 1737 &  6 & Sdm&   N5746   \\
PGC052534 & J144229.6+013001& 15.21 &         & 1876 & 27 & S0 &   N5746   \\
NGC5733   & J144245.9-002104& 14.77 & 25.9 TF & 1655 & 16 & Scd&   N5746   \\
PGC1216689& J144246.0+020725& 16.85 &         & 1549 &  4 & Sdm&   N5746   \\
UGC09482  & J144247.0+003942& 15.58 & 36.1 TF & 1782 & 10 & Scd&   N5746   \\
PGC1141860& J144247.7-002954& 17.79 &         & 1718 &  4 & Im &   N5746   \\
IC1048    & J144258.0+045323& 13.95 & 29.4 tf & 1603 &  6 & S0 &   IC1048  \\
PGC135860 & J144300.2-002300& 16.61 &         & 1688 & 20 & Ir &           \\
UGC09485  & J144302.8+044555& 15.4  & 31.3 TF & 1672 &  7 & Scd&   IC1048  \\
AGC242618 & J144329.2+043153& 16.65 &         & 1710 & 17 & Sdm&   IC1048  \\
SDSSJ14434& J144346.9+033234& 17.98 &         & 1620 &  4 & BCD&   N5746   \\
NGC5738   & J144356.4+013615& 14.77 &         & 1700 & 21 & S0 &   N5746   \\
NGC5740   & J144424.4+014047& 12.60 & 29.0 tf & 1526 &  6 & Sb &   N5746   \\
PGC052652 & J144430.9+013122& 15.69 &         & 1428 & 20 & Im &   N5746   \\
NGC5746   & J144456.0+015717& 11.34 & 29.0 tf & 1680 & 10 & Sb &   N5746   \\
PGC1150429& J144515.8-000934& 15.74 &         & 1598 &  7 & Ir &   N5746   \\
UGC09500  & J144521.4+075145& 16.0  & 19.1 TF & 1671 &  4 & Sm &           \\
NGC5750   & J144611.1-001323& 12.55 & 32.4 tf & 1635 & 10 & S0a&   N5746   \\
AGC242625 & J144620.2+044359& 17.93 & 20.0 TF & 1491 & 17 & Im &   IC1048  \\
AGC245015 & J144834.4+052553& 17.16 &         & 1672 & 17 & Ir &           \\
PGC3277346& J144835.2-041758& 17.48 &         & 1763 & 89 & Ir &           \\
PGC052893 & J144848.0-034259& 15.06 & 17.3 TF &  868 &  6 & Sdm&           \\
AGC249197 & J144950.7+095630& 18.69 &         & 1802 & 40 & BCD&           \\
PGC1241857& J145022.9+025729& 17.18 &         & 1659 &  3 & BCD&   N5775   \\
SDSSJ14505& J145059.9+022016& 18.56 &         & 1516 & 46 & dE &   N5746   \\
SDSSJ14510& J145106.8+023127& 18.53 &         & 2057 & 21 & Im &           \\
PGC1350028& J145132.8+083600& 17.68 &         & 1758 &  1 & dEn&           \\
SDSSJ14520& J145201.9+025842& 17.86 &         & 1733 &239 & BCD&   N5775   \\
NGC5768   & J145207.9-023147& 13.52 & 28.8 tf & 1906 &  4 & Sc &           \\
VV 815    & J145234.8-033342& 15.14 & 27.3 TF & 1893 &  5 & Sdm&           \\
AGC249264 & J145243.3+043617& 18.4  & 22.9 TF & 1563 & 17 & Ir &           \\
VV 815    & J145254.9-034936& 15.12 & 22.2 TF & 1805 & 30 & Sdm&           \\
IC1066    & J145302.9+031746& 14.27 &         & 1546 &  6 & Sb &   N5775   \\
IC1067    & J145305.2+031954& 13.62 &         & 1546 &  6 & Sb &   N5775   \\
NGC5770   & J145315.0+035735& 13.18 & 19.0 sbf& 1462 & 14 & S0a&   N5775   \\
NGC5774   & J145342.5+033457& 13.10 & 20.4 TF & 1526 &  3 & Sd &   N5775   \\
NGC5775   & J145357.6+033239& 12.23 & 21.4 tf & 1652 &  5 & Sb &   N5775   \\
KKR 14    & J145443.0+010943& 17.80 &         & 1791 &  7 & Ir &   N5846   \\
PGC1197564& J145455.8+013132& 18.77 &         & 1730 & 18 & Ir &   N5846   \\
PGC1184577& J145509.3+010602& 18.35 &         & 1664 &  5 & Ir &   N5846   \\
PGC053365 & J145542.9-010032& 15.78 &         & 1802 &  3 & Sm &   N5792   \\
UGC09601  & J145601.8-012316& 14.62 &         & 1791 &  6 & Scd&   N5792   \\
PGC1083529& J145620.3-024542& 18.66 &         & 1963 & 81 & Ir &           \\
PGC1186917& J145634.3+011045& 17.24 &         & 1904 & 23 & dS0&   N5846   \\
PGC184824 & J145744.8-025913& 17.14 &         & 1736 & 55 & dEn&           \\
PGC1179522& J145753.1+005603& 16.90 &         & 1847 & 16 & S0 &   N5846   \\
PGC184842 & J145807.8-011845& 16.45 &         & 1910 & 47 & S0 &   N5792   \\
AGC736339 & J145808.8+064448& 18.23 & 20.7 TF & 1621 & 15 & Ir &           \\
PGC184851 & J145821.2+015043& 15.99 &         & 1858 & 22 & E  &   N5846   \\
NGC5792   & J145822.6-010527& 12.12 & 22.5 tf & 1878 &  5 & Sb &   N5792   \\
SDSSJ14582& J145828.6+013235& 17.64 &         & 1461 & 20 & Sph&   N5811   \\
CGCG48-085& J145837.8+064630& 15.4  & 27.4 TF & 1667 & 12 & Sm &           \\
PGC1223766& J145840.9+022024& 18.36 &         & 1515 & 21 & dEn&   N5846   \\
PGC1242097& J145846.1+025808& 16.39 &         & 1776 & 49 & BCD&   N5846   \\
PGC053521 & J145848.7+020125& 14.87 &         & 1779 & 12 & E  &   N5846   \\
SDSSJ14594& J145944.8+020752& 18.39 &         & 1432 & 59 & Sph&           \\
NGC5806   & J150000.4+015329& 12.35 & 25.6 tf & 1328 &  5 & Sb &   N5838   \\
PGC053577 & J150001.3-010528& 15.81 &         & 1842 &  3 & BCD&   N5792   \\
PGC053587 & J150016.6+021802& 15.50 &         & 1800 & 17 & SO &   N5846   \\
SDSSJ15001& J150019.2+005700& 17.56 &         & 1927 &  2 & Sph&           \\
NGC5811   & J150027.0+013725& 14.76 &         & 1498 & 16 & Sm &   N5811   \\
SDSSJ15003& J150033.0+021349& 17.24 &         & 1237 & 36 & Sph&   N5838   \\
PGC1193898& J150052.6+012418& 16.87 &         & 1868 & 34 & dEn&   N5846   \\
SDSSJ15010& J150100.9+010050& 18.03 &         & 1716 & 21 & Ir &   N5846   \\
MRK 1390  & J150103.1+004227& 15.88 &         & 1722 & 15 & dS0&   N5846   \\
SDSSJ15010& J150107.0+020525& 18.33 &         & 1921 & 26 & Sph&   N5846   \\
NGC5813   & J150111.3+014207& 11.52 & 32.1 sbf& 1942 &  6 & E  &   N5846   \\
PGC1205406& J150115.9+014625& 18.06 &         & 1345 & 44 & dE &   N5838   \\
UGC09661  & J150203.5+015028& 14.81 &         & 1213 &  5 & Sdm&   N5838   \\
PGC1192611& J150228.2+012151& 18.49 &         & 1500 & 44 & Sph&   N5846   \\
SDSSJ15023& J150233.0+015609& 18.18 &         & 1619 & 60 & Sph&   N5846   \\
SDSSJ15023& J150236.0+020139& 18.11 &         & 1946 & 29 & dE &   N5846   \\
PGC1230503& J150344.2+023309& 17.56 &         & 1743 & 27 & dE &   N5846   \\
SDSSJ15034& J150349.9+005835& 16.99 &         & 1971 & 80 & Im &           \\
PGC1185375& J150350.3+010737& 16.48 &         & 1539 & 19 & dE &   N5846   \\
KKR 15    & J150356.2+002546& 17.32 & 23.7 TF & 1555 &  6 & BCD&   N5846   \\
NGC5831   & J150407.0+011312& 12.44 & 27.5 sbf& 1626 &  5 & E  &   N5846   \\
PGC1197513& J150408.4+013128& 16.43 &         & 1816 & 17 & S0a&   N5846   \\
PGC1230189& J150413.1+023235& 15.89 &         & 1887 & 22 & E  &   N5846   \\
PGC1179083& J150423.8+005506& 18.12 &         & 1626 &  2 & dE &           \\
PGC1216386& J150424.7+020653& 17.44 &         & 1674 & 23 & dE &   N5846   \\
UGC09682  & J150430.2-005105& 14.9  & 28.7 TF & 1777 &  7 & Sm &   N5792   \\
KKR 16    & J150434.2-023513& 16.51 & 27.3 TF & 1571 &  4 & Ir &           \\
PGC1190315& J150442.9+011727& 16.96 &         & 1944 & 24 & dE &   N5846   \\
SDSSJ15044& J150448.5+015851& 18.07 &         & 1937 & 23 & Sph&   N5846   \\
NGC5838   & J150526.3+020558& 11.79 & 27.0 tf & 1334 & 10 & S0 &   N5838   \\
NGC5839   & J150527.5+013805& 13.69 & 22.6 sbf& 1198 & 15 & S0 &   N5838   \\
PGC1199471& J150531.8+013516& 18.15 &         &  947 & 35 & dEn&           \\
PGC1190714& J150537.7+011811& 17.43 &         & 2058 & 22 & dEn&   N5846   \\
PGC1209872& J150550.5+015430& 16.93 &         & 1701 & 26 & dE &   N5846   \\
PGC1213020& J150553.2+020028& 18.35 &         & 1266 & 20 & Ir &   N5838   \\
PGC3092767& J150559.5-042741& 16.32 &         & 1916 & 45 & Sa &           \\
NGC5845   & J150600.8+013802& 13.44 & 25.9 sbf& 1424 & 10 & E  &   N5846   \\
PGC1218738& J150603.4+021106& 16.34 &         & 1648 & 23 & Sm &   N5846   \\
PGC1215798& J150611.3+020545& 17.64 &         & 1787 &  3 & Sd &   N5846   \\
NGC5846   & J150629.2+013623& 11.09 & 25.6 sbf& 1688 &  5 & E  &   N5846   \\
SDSSJ15063& J150634.2+001256& 17.87 &         & 1979 & 90 & dEn&   N5869   \\
PGC3119319& J150634.2+013332& 16.13 &         & 1488 & 24 & E  &   N5846   \\
NGC5841   & J150635.0+020018& 14.55 &         & 1220 & 15 & S0 &   N5838   \\
PGC1156476& J150641.0+000436& 18.07 &         & 1678 & 30 & dE &   N5846   \\
SDSSJ15065& J150658.3+015940& 17.8  &         & 1283 &  1 & Sph&   N5838   \\
PGC1067957& J150704.8-034201& 16.39 &         & 1638 & 45 & Sm &           \\
PGC1085904& J150708.1-023946& 17.92 &         & 1992 &  4 & Im &           \\
PGC1185172& J150734.2+010714& 17.64 &         & 1606 & 37 & dE &   N5846   \\
PGC054004 & J150737.2+020110& 15.86 &         & 1907 & 30 & dEn&   N5846   \\
NGC5854   & J150747.7+023407& 12.65 & 23.5 TF & 1716 &  8 & S0 &   N5846   \\
PGC054016 & J150747.8+011732& 15.67 &         & 2073 & 24 & dEn&   N5869   \\
PGC1217593& J150801.4+020904& 18.04 &         & 1048 & 20 & dE &           \\
PGC054037 & J150805.6+013906& 16.08 &         & 1813 & 20 & Sa &   N5846   \\
NGC5846:[M& J150808.5+025418& 18.2  &         & 1663 &  1 & Ir &           \\
SDSSJ15081& J150812.4+012959& 18.02 &         & 1579 & 35 & Sph&   N5846   \\
PGC1206166& J150822.7+014755& 18.08 &         & 1668 & 39 & Sph&   N5846   \\
NGC5846:[M& J150825.6+014225& 18.9  &         & 2025 &  1 & Sph&           \\
PGC1209573& J150847.2+015400& 16.67 &         & 1989 & 21 & dE &   N5846   \\
PGC1176385& J150904.3+004919& 16.81 &         & 1616 &  3 & Im &   N5846   \\
SDSSJ15090& J150907.9+004329& 17.69 &         & 1642 & 23 & Ir &   N5846   \\
PGC1210284& J150914.9+015517& 16.64 &         & 1717 & 17 & dE &   N5846   \\
PGC1128787& J150933.6-010118& 17.56 &         & 1820 &  4 & En &   N5846   \\
NGC5864   & J150933.6+030310& 12.70 & 27.0 tf & 1868 &  4 & S0 &   N5846   \\
NGC5869   & J150949.4+002812& 13.15 & 24.9 sbf& 2058 &  4 & E  &   N5869   \\
UGC09746  & J151016.8+015601& 14.84 & 30.4 TF & 1714 &  6 & Sbc&   N5846   \\
UGC09751  & J151058.5+012615& 15.62 & 27.0 TF & 1551 & 15 & Scd&   N5846   \\
PGC1202458& J151101.3+014050& 17.28 &         & 1653 & 23 & dE &   N5846   \\
SDSSJ15112& J151121.5+013639& 17.8  &         & 1900 & 76 & Sph&           \\
UGC09760  & J151202.5+014155& 15.20 & 22.4 TF & 2003 &  2 & Sd &   N5846   \\
PGC1199418& J151208.2+013509& 17.00 &         & 1937 & 17 & En &   N5846   \\
PGC1215336& J151224.0+020448& 16.94 &         & 1675 & 24 & dEn&   N5846   \\
PGC1176138& J151231.7+004845& 16.32 &         & 1806 & 60 & Ir &   N5846   \\
PGC1200646& J151241.4+013724& 17.25 &         & 1863 & 36 & dE &   N5846   \\
AGC258278 & J151245.0+060951& 18.2  & 23.3 TF & 1485 & 22 & Im &           \\
PGC1236445& J151509.4+024507& 16.89 &         & 1757 & 21 & dS0&   N5846   \\
PGC054452 & J151534.6+021454& 14.82 &         & 1834 & 22 & S0 &           \\
UGC09787  & J151542.8+012721& 15.40 & 33.0 TF & 1576 & 10 & Scd&           \\
PGC1234821& J151606.0+024131& 17.23 &         & 1448 &  1 & Sd &           \\
PGC3124577& J151729.4+033508& 17.36 &         & 1892 & 19 & BCD&           \\
PGC1230249& J151913.4+023242& 16.17 &         & 1873 &  1 & S0 &           \\
AGC252596 & J151947.9+035841& 17.7  &         & 1708 & 56 & Scd&           \\
NGC5913   & J152055.4-023441& 14.02 & 32.0 tf & 1975 & 11 & Sa &           \\
NGC5921   & J152156.5+050413& 11.68 & 20.6 sn & 1483 &  1 & Sbc&           \\
UGC09830  & J152300.8+043145& 15.91 & 36.5 tf & 1831 &  5 & Sc &           \\
PGC054944 & J152320.1-040906& 13.51 & 29.4 tf & 1975 &  6 & Sab&           \\
AGC258405 & J152411.4+072918& 17.25 & 37.8 TF & 1766 &  1 & Sd &           \\
AGC258127 & J152432.9+083056& 17.4  &         & 1865 & 42 & Sm &           \\
PGC3123131& J152450.1+030453& 17.58 &         & 1744 &  2 & BCD&           \\
UGC09845  & J152605.5+091215& 15.43 & 32.6 tf & 1918 &  2 & Sbc&           \\
SDSSJ15265& J152655.4+094657& 18.23 &         & 1884 & 71 & Sph&           \\
AGC251585 & J152744.5+094157& 16.93 &         & 1840 & 20 & Sm &           \\
SDSSJ15284& J152845.0+042421& 17.16 &         & 1793 &  0 & BCD&           \\
PGC3123706& J153016.7+031722& 17.62 &         & 1798 &  3 & BCD&           \\
KKR 20    & J153106.1+034947& 16.50 & 31.2 TF & 1781 &  2 & Ir &           \\
SDSSJ15313& J153133.3+060131& 18.25 &         & 1947 &  1 & Im &           \\
PGC1164369& J153536.1+002231& 17.68 &         & 2022 &  4 & Im &           \\
NGC5964   & J153736.2+055825& 13.42 & 27.3 tf & 1468 &  2 & Sd &           \\
AGC258331 & J154106.3+044510& 18.04 &         & 1996 &  8 & Im &           \\
AGC258332 & J154156.9+043448& 18.0  & 27.0 TF & 1992 & 26 & Ir &           \\
UGC09977  & J154159.6+004246& 14.61 & 28.9 tf & 1918 &  3 & Sbc&   U9977   \\
UGC09979  & J154219.4+002829& 14.94 & 22.4 TF & 1963 &  3 & Ir &   U9977   \\
UGC10023  & J154609.7+065354& 15.66 & 29.0 TF & 1440 &  5 & Sm &           \\
UGC10025  & J154624.3+025039& 16.88 & 32.9 TF & 1538 &  2 & Sd &           \\
SDSSJ15462& J154624.6+042921& 18.37 &         & 1879 &  1 & BCD&           \\
AGC258343 & J154958.0+053102& 18.44 & 27.5 TF & 1979 & 29 & Ir &           \\
NGC6010   & J155419.2+003235& 13.14 & 21.6 tf & 1903 & 11 & S0 &           \\
SDSSJ15552& J155522.4+025515& 18.90 &         & 2040 & 21 & Ir &           \\
SDSSJ15561& J155614.4+060553& 17.6  &         & 1857 & 46 & Ir &   N6017   \\
AGC252891 & J155636.8+061139& 16.99 & 32.2 TF & 1957 &  5 & Ir &           \\
AGC258349 & J155637.0+055807& 17.5  & 29.5 TF & 1900 & 37 & Ir &           \\
NGC6017   & J155715.4+055954& 13.80 & 29.8 sbf& 1835 &  8 & S0 &   N6017   \\
AGC258430 & J155955.5+065303& 18.12 & 27.7 TF & 1541 & 20 & Ir &           \\
IC1158    & J160134.1+014228& 13.3  & 30.5 tf & 1953 &  6 & Sc &           \\
AGC716496 & J160148.4+065018& 17.9  &         & 1820 &  3 & Ir &           \\
SDSSJ16024& J160247.9+065231& 18.0  &         & 1516 &  3 & BCD&           \\
KKSG 48   & J160540.8-043419& 17.5  & 34.9 TF & 1615 &  1 & Sm &           \\
CGCG079-46& J160602.1+083024& 15.8  & 37.8 TF & 1420 & 14 & Sm &           \\
CGCG051-43& J160641.0+063451& 15.2  & 20.4 TF & 1805 & 24 & Im &           \\
IC1197    & J160817.2+073218& 14.1  & 25.3 tf & 1424 &  4 & Sd &           \\
PGC057323 & J160936.8-043716& 15.5  & 10.7 tf &  838 &  6 & Im &           \\
UGC10229  & J160943.9-000655& 16.6  &         & 1522 &  5 & Im &           \\
NGC6070   & J160958.7+004234& 12.45 & 31.3 tf & 2033 &  6 & Scd&   N6070   \\
AGC268216 & J161220.5+063237& 17.7  & 23.5 TF & 1778 & 23 & Ir &           \\
UGC10290  & J161433.0+004918& 15.2  &         & 2017 &  4 & Sm &   N6070   \\
NGC6106   & J161847.1+072436& 12.84 & 24.6 tf & 1512 &  7 & Sc &           \\
CGCG024-01& J161918.2+014211& 15.4  & 29.9 TF & 1565 &  7 & Sm &           \\
NGC6118   & J162148.6-021700& 11.0  & 23.4 tf & 1601 &  2 & Sc &           \\
CGCG052-15& J162604.3+025424& 15.2  &         & 1599 & 32 & S0 &           \\
PGC058661 & J163808.6-044918& 15.5  & 26.5 tf & 1612 &  5 & Sd &   P58798  \\
PGC058798 & J164203.1-050157& 14.87 & 25.3 tf & 1600 &  2 & Sbc&   P58798  \\
UGC10554  & J164821.3-013708& 14.90 & 32.0 TF & 1617 &  9 & Scd&           \\
KKR 30    & J165638.5+075955& 17.0  & 30.7 TF & 1584 &  6 & Ir &           \\
UGC10862  & J172809.0+072521& 14.4  & 26.6 TF & 1816 &  8 & Sc &   N6384   \\
NGC6384   & J173224.2+070337& 11.14 & 23.9 tf & 1792 &  8 & Sbc&   N6384   \\
UGC11030  & J175434.2+025251& 14.9  & 20.1 TF & 1884 &  9 & Sc &           \\
UGC11074  & J175907.4+070829& 13.9  & 25.7 tf & 2044 &  9 & Scd&           \\
NGC6509   & J175925.3+061713& 13.1  & 34.6 TF & 1957 &  9 & Sc &           \\
\end{longtable}
\end{center}

\clearpage
\begin{center}
\begin{table}
\caption{Groups in the Bootes strip}
\begin{tabular}{lrrcrrrccrr} \hline
Group  & $N_v$& $<V_{LG}>$&  $D$  & $\sigma_v$ & $R_h$ &  $\log M_*$ & $\log M_p$ & $\log M_p/M_*$&  $N_D$ & $\sigma(m-M)$ \\
\hline
&   &  km/s & Mpc &  km/s & kpc & $M_\odot$  & $M_\odot$ &        &      &          \\
\hline
N4900  &  8 &  779 & 23.6&   36  & 116 & 10.67 & 11.95&  1.28  &   1  & (0.4)    \\
	&    &      &     &       &     &       &      &        &      &          \\
N5248  &  5 & 1116 & 17.1&   38  & 151 & 10.88 & 12.06&  1.18  &   4  &  0.43    \\
	&    &      &     &       &     &       &      &        &      &          \\
N5363  & 17 & 1173 & 17.5&  144  & 165 & 11.25 & 12.80&  1.55  &   9  &  0.41    \\
	&    &      &     &       &     &       &      &        &      &          \\
N5506  &  4 & 1777 & 23.8&   23  &  35 & 10.94 & 11.07&  0.13  &   1  & (0.4)    \\
	&    &      &     &       &     &       &      &        &      &          \\
N5566  & 12 & 1542 & 22.3&  103  & 196 & 11.35 & 12.81&  1.46  &   6  &  0.19    \\
	&    &      &     &       &     &       &      &        &      &          \\
N5638  & 12 & 1684 & 25.8&   74  & 203 & 10.90 & 12.21&  1.31  &   5  &  0.44    \\
	&    &      &     &       &     &       &      &        &      &          \\
P51971 &  3 & 1483 & 20.6&   10  & 100 &  9.04 & 10.35&  1.31  &   0  &   -      \\
	&    &      &     &       &     &       &      &        &      &          \\
IC1048 &  4 & 1619 & 26.4&   83  & 150 & 10.38 & 12.32&  1.94  &   3  &  0.43    \\
	&    &      &     &       &     &       &      &        &      &          \\
N5746  & 38 & 1671 & 25.6&  107  & 296 & 11.74 & 13.24&  1.50  &  15  &  0.40    \\
	&    &      &     &       &     &       &      &        &      &          \\
N5775  &  7 & 1589 & 20.1&   87  & 120 & 11.01 & 12.66&  1.65  &   3  &  0.10    \\
	&    &      &     &       &     &       &      &        &      &          \\
N5792  &  6 & 1833 & 25.4&   48  & 290 & 11.06 & 12.00&  0.94  &   2  &  0.26    \\
	&    &      &     &       &     &       &      &        &      &          \\
N5838  &  9  &1269 & 25.0&   53  & 210 & 11.31 & 12.29&  0.98  &   3  &  0.16    \\
	&     &     &     &       &     &       &      &        &      &          \\
N5846  & 74  &1803 & 26.4&  228  & 415 & 11.84 & 13.66&  1.82  &   9  &  0.24    \\
\hline
\end{tabular}
\end{table}
\end{center}

\begin{center}
\begin{table}
\caption{Pairs in the Bootes strip}
\begin{tabular}{lclcrrcccr} \hline
Name    & $ <V_{LG}>$ & $\Delta V_{12}$ &  $D$   &  $R_{12}$ & $\log M_*$&  $M_p$   &$log M_p/M_*$ & $N_D$& $\sigma(m-M)$ \\
\hline
&    km/s &  km/s & Mpc  &  kpc &$M_\odot$ &$M_\odot$  &        &    &   mag    \\
\hline
P1078534 &  1134   &21$\pm$ 9 & 29.4 &  68  & 8.96& 10.55 & 1.59   &  2 &   0.04   \\
P1080976 &         &       &      &      &     &       &        &    &          \\
U8995    &  1168   &45$\pm$ 8 & 23.6 &  12  & 9.59& 10.45 & 0.86   &  1 &  (0.4)   \\
P2807027 &         &       &      &      &     &       &        &    &          \\
N5496    &  1446   &19$\pm$17 & 24.1 & 305  &10.24& 11.11 & 0.85   &  1 &  (0.4)   \\
IC976    &         &       &      &      &     &       &        &    &          \\
U9299    &  1466   &14$\pm$20 & 28.3 & 304  & 9.84& 10.84 & 1.00   &  2 &   0.18   \\
P051719  &         &       &      &      &     &       &        &    &          \\
N5669    &  1361   &24$\pm$ 6 & 19.7 &  35  & 9.81& 10.37 & 0.56   &  1 &  (0.4)   \\
CGCG75-63&         &       &      &      &     &       &        &    &          \\
N5811    &  1480   &37$\pm$26 & 20.6 & 180  & 9.34& 11.46 & 2.12   &  - &    -     \\
SDSS14582&         &       &      &      &     &       &        &    &          \\
U9977    &  1940   &45$\pm$ 4 & 25.4 & 111  &10.14& 11.42 & 1.28   &  2 &   0.28   \\
U9979    &         &       &      &      &     &       &        &    &          \\
N6017    &  1846   &22$\pm$47 & 29.8 & 141  &10.34& 10.90 & 0.56   &  1 &  (0.4)   \\
SDSS15561&         &       &      &      &     &       &        &    &          \\
N6070    &  2025   &16$\pm$ 7 & 31.3 & 628  &10.87& 11.28 & 0.41   &  1 &  (0.4)   \\
U10290   &         &       &      &      &     &       &        &    &          \\
P58798   &  1606   &12$\pm$6 & 25.9 & 450  &10.60& 10.88 & 0.28   &  2 &   0.05   \\
P58661   &         &       &      &      &     &       &        &    &          \\
N6384    &  1804   &24$\pm$11 & 25.1 & 488  &11.12& 11.52 & 0.40   &  2 &   0.12   \\
U10862   &         &       &      &      &     &       &        &    &          \\
\hline
\end{tabular}
\end{table}
\end{center}

\end{document}